\documentclass{article}

\usepackage[main,final,nonatbib]{neurips_2026}
\usepackage{booktabs}

\usepackage[utf8]{inputenc}
\usepackage[T1]{fontenc}
\usepackage{hyperref}
\usepackage{url}
\usepackage{booktabs}
\usepackage{amsfonts}
\usepackage{nicefrac}
\usepackage{microtype}
\usepackage{xcolor}

\usepackage{amsmath}
\usepackage{graphicx}

\usepackage[numbers,sort&compress]{natbib}
\usepackage{subcaption}
\usepackage{makecell}
\usepackage{placeins}
\usepackage{enumitem}

\title{CellMSA: Context Modeling for \\
Single-Cell Representation Learning}

\author{%
  \textbf{Suyuan Zhao}$^{1,3}$\thanks{Equal contribution.} \quad
  \textbf{Minghao Liu}$^{1,4,5}$\footnotemark[1] \quad
  \textbf{Yizhen Luo}$^{1,3}$ \quad
  \textbf{Zaiqing Nie}$^{1,2}$\thanks{Corresponding author.} \\
  $^{1}$Institute for AI Industry Research (AIR), Tsinghua University \\
  $^{2}$PharMolix Inc. \\
  $^{3}$Department of Computer Science and Technology, Tsinghua University \\
  $^{4}$Tsinghua Institute of Multidisciplinary Biomedical Research (TIMBR), Tsinghua University \\
  $^{5}$National Institute of Biological Sciences (NIBS) \\
  \texttt{\{zhaosy23,liumh23,yz-luo22\}@mails.tsinghua.edu.cn} \\
  \texttt{zaiqing@air.tsinghua.edu.cn}
}

\begin{document}

\maketitle

\begin{abstract}
Single-cell transcriptomics enables profiling of cellular states at unprecedented resolution, but its high dimensionality, sparsity, and technical batch effects pose significant challenges for representation learning.
Existing single-cell foundation models typically encode each cell independently or only model cells from the same batch for denoising, thereby underutilizing the rich relational information across batches and cell types to model gene expression patterns.
We argue that single-cell models can benefit from more informative cell-context modeling. By comparing consistency and variation across cells, models can capture fine-grained gene-gene dependencies associated with cell states, which are essential for learning high-quality representations.
Inspired by the use of multiple sequence alignment (MSA) context in protein modeling, we propose \textbf{CellMSA}, a single-cell representation learning framework that introduces an MSA-inspired inductive bias into transcriptomic modeling. For each target cell, CellMSA retrieves relevant cells from different batches and biologically related cell types as context, and summarizes cross-cell patterns into a context-dependent gene-pair representation. This representation is then injected into a pair-aware target-cell encoder for fine-grained representation learning.
We pretrain CellMSA on a large-scale human single-cell corpus of approximately 109 million cell observations, including 65.6 million primary observations. Experiments show that our framework consistently outperforms existing methods across multiple benchmarks. Code is available at the following repository: \url{https://github.com/PharMolix/CellMSA}.
\end{abstract}

\section{Introduction}

The development of single-cell RNA sequencing (scRNA-seq) has enabled researchers to characterize cellular state transitions and disease-associated heterogeneity at single-cell resolution \cite{tang2009mrna, zheng2017massively, ziegenhain2017comparative}, and has in turn driven rapid progress in single-cell foundation models. 
In recent years, large-scale pretraining has made it possible to learn transferable representations of cells from massive single-cell transcriptomic datasets, showing promise across a range of downstream tasks, including cell type annotation, batch integration, and perturbation modeling \cite{ScBERT_yang_2022, cui2024scgpt, geneformer, scfoundation, zhao2024langcell, rosen2023universal, pearce2025cross}.
However, most existing single-cell foundation models treat each cell as an isolated input for modeling. This paradigm has inherent limitations: single-cell transcriptomes are highly sparse and are strongly affected by sequencing depth, dropout, and batch effects \cite{hicks2018missing, lahnemann2020eleven}. As a result, when the model relies on only a single observation of a cell, it is often difficult to distinguish biological variation from stochastic noise, or to recover fine-grained gene dependency patterns associated with a given cell state \cite{kedzierska2025zero, ahlmann2025deep}.

Recent models begin to incorporate multi-cell context explicitly. By jointly modeling multiple cells from the same sequencing batch, these methods have improved over earlier single-cell-only approaches \cite{wen2023cellplm, adduri2025predicting, dong2026stack}. However, this line of work still has important limitations. 
First, to simultaneously model multiple cells with large information content, existing methods typically compress single-cell gene expression profiles using strategies such as gene-weighted averaging \cite{wen2023cellplm}, cell encoders \cite{adduri2025predicting}, or gene modules \cite{dong2026stack}. Such compression can discard gene-level information and therefore limit the model’s ability to capture fine-grained gene-gene dependencies. 
Second, existing approaches often organize together only cells of the same type from the same sequencing batch. Such context mainly reflects shared patterns under local experimental conditions and therefore provides limited additional information beyond denoising. Because they do not explicitly exploit cross-batch signals or biologically related cell types, these methods may be less effective at improving robustness across batches or identifying condition-specific gene co-expression patterns.

\begin{figure}
    \centering
    \includegraphics[width=0.9\linewidth]{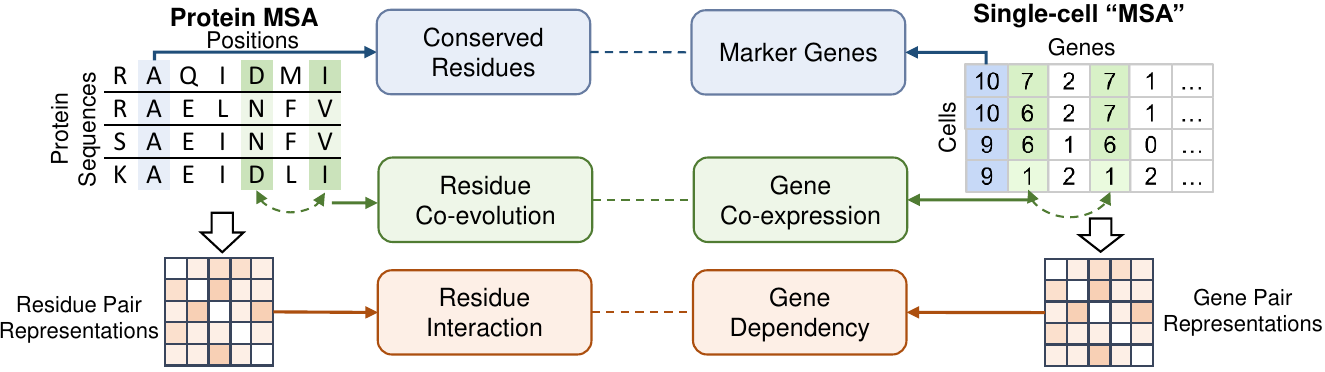}
    \caption{\textbf{Utilizing MSA context to enhance protein modeling has a natural counterpart in single-cell modeling.} Protein models capture \textit{\textbf{conserved residues}} closely related to structural stability, as well as residue \textit{\textbf{co-evolution}}, thereby modeling structure-aware \textit{\textbf{residue-pair representations}}. Similarly, CellMSA learns the \textit{\textbf{marker genes}} and gene \textit{\textbf{co-expression}} patterns from context, and in turn models cell-state-dependent \textit{\textbf{gene dependencies}}.}
    \label{fig1}
    \vskip -0.25in
\end{figure}

These observations suggest that single-cell foundation models require a new inductive bias that enables the use of more informative cellular context while maintaining gene-level resolution.
To this end, we propose CellMSA, inspired by the use of multiple sequence alignment (MSA) context in AlphaFold-style protein structure modeling~\cite{jumper2021highly, abramson2024accurate}. 
Protein MSA context enables AlphaFold models to construct residue-pair representations by aggregating conservation and co-evolution across homologous sequences, thereby recovering stable structural constraints.
We argue that a similar principle can be transferred to single-cell modeling (as shown in Figure~\ref{fig1}). By aggregating consistency and variation across similar cells, CellMSA can identify the marker genes of the cell state and capture gene co-expression patterns. 
Motivated by the principle, CellMSA retrieves cells from multiple batches and similar cell types, and distills this cross-cell evidence into a gene-pair representation, which is used to guide target-cell encoding for more robust and refined representation learning.

Specifically, for each target cell, we retrieve a set of similar cells, including cells from the same batch, cells of the same type from different batches, and cells from biologically related cell types, and organize them as an MSA-like context. 
We first introduce a CellMSA-Module that uses low-dimensional gene-level embeddings to compare consistent and differential expression patterns between the target cell and the retrieved cells, and integrate this cross-cell evidence into a gene-pair representation. We then introduce a GenePairformer module to inject the updated gene-pair representation into target-cell encoding. 
We designed three single-cell-specific pretraining objectives and pretrained CellMSA on a large-scale corpus containing approximately 109 million human cell observations, including 65.6 million primary observations \cite{cellxgeneDiscover}.

To validate CellMSA’s ability to effectively leverage contextual information to enhance cell representation learning, we established a comprehensive benchmark covering multiple important biological tasks.
Specifically, CellMSA outperforms the strongest baseline within the label-informed batch integration setting \cite{scib, Tabula_Sapiens} and achieves the highest macro-F1 scores for cell type and cell state classification \cite{Tabula_Sapiens, de2021rationale}, including comparisons with CellPLM. It also improves Pearson $\Delta$ for perturbation prediction by 8.8\% over the strongest baseline \cite{adduri2025predicting, replogle2022mapping}. An additional label-free integration experiment evaluates performance without cell-type annotations for context retrieval.
We further provide additional qualitative and quantitative results to intuitively demonstrate that the model effectively captures gene relationships by aligning cellular contexts, highlighting the interpretability of CellMSA.

In summary, our main contributions are as follows:

\begin{itemize}[noitemsep,topsep=0pt,parsep=5pt,partopsep=0pt,leftmargin=15pt]
\item We propose CellMSA, a framework that introduces an MSA-inspired inductive bias by modeling gene dependencies from informative context to enhance single-cell representation learning.

\item We perform large-scale pretraining of CellMSA on a corpus containing approximately 109 million human cell observations, including 65.6 million primary observations, by jointly optimizing three single-cell-specific objectives.

\item We demonstrate the outstanding performance of CellMSA across a variety of downstream tasks related to single-cell analysis.
\end{itemize}

\section{Related work}

\paragraph{Single-cell transcriptomics foundation models.}
In recent years, single-cell foundation models have advanced rapidly. Representative methods such as Geneformer \cite{geneformer}, scGPT \cite{cui2024scgpt}, scFoundation \cite{scfoundation} and UCE \cite{rosen2023universal} rely on large-scale pretraining to learn transferable representations of cells and genes, and have been applied to tasks including cell type annotation, batch integration, and perturbation modeling. Collectively, these studies demonstrate the promise of large-scale pretraining on single-cell data. However, most existing approaches still follow the paradigm of independent single-cell encoding, which remains limited in robustness and the ability to capture fine-grained gene dependencies.

\paragraph{Methods incorporating multi-cell context.}
To address the limitations of independent single-cell modeling, recent studies have begun to explicitly introduce \emph{multi-cell context}. CellPLM \cite{wen2023cellplm} was among the earlier works to propose pretraining with ``cells as tokens'' and ``tissues as sentences''. Recently, STATE \cite{adduri2025predicting} has utilized cell representations from multiple cohorts for perturbation prediction tasks. Stack \cite{dong2026stack} further adopts an in-context learning framework, organizing multiple cells as gene module blocks for joint input. 
Compared with these methods, CellMSA constructs an MSA-like context from same-batch, cross-batch, and biologically related cells, while preserving gene-level alignment across cells to support context-dependent gene-pair modeling.

\paragraph{MSA-enhanced modeling in proteins.}
In protein modeling, multiple sequence alignment (MSA) serves as a key source of information for identifying conserved sites and recovering residue relationships from coordinated variation \cite{jumper2021highly, abramson2024accurate, zhang2021co, ju2021copulanet, rao2021msa, marks2011protein, morcos2011direct}. 
Early studies show that residue co-variation patterns extracted from MSAs can be used to infer residue conservation signals and inter-residue relationships \cite{marks2011protein, morcos2011direct}. Representatively, the AlphaFold series \cite{jumper2021highly, abramson2024accurate} further integrates MSA representations with pair representations in deep neural architectures, enabling the model to extract residue-pair relationships from multi-sequence context.
Inspired by this line of work, we transfer the modeling logic of recovering higher-order relational structure from consistency and variation across similar samples to the single-cell scenario.

\section{Methods}

\begin{figure}[t]
    \centering
    \includegraphics[width=1\linewidth]{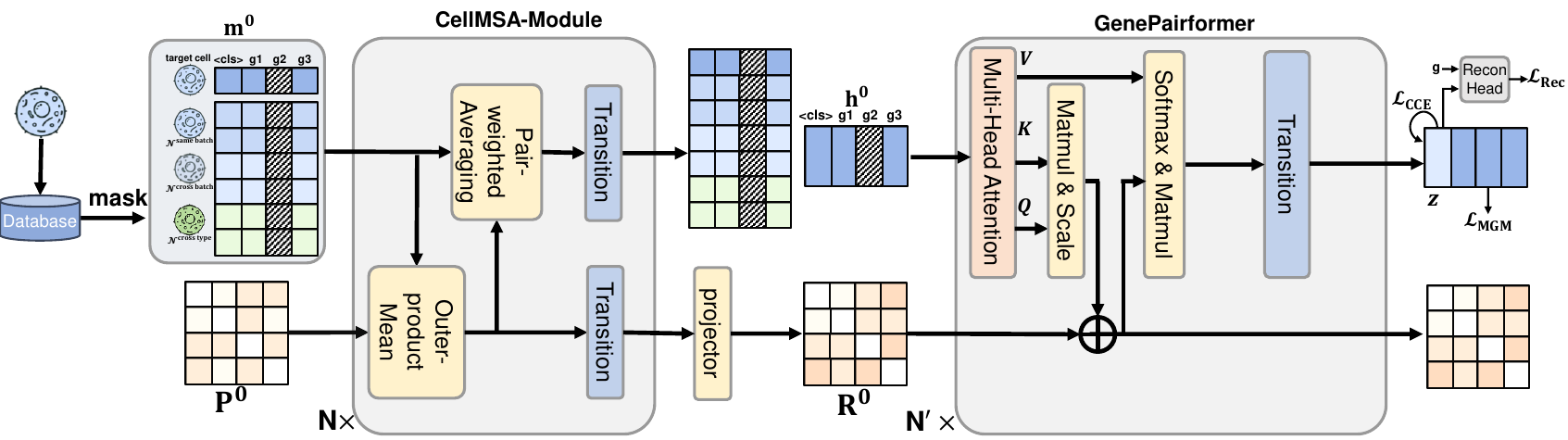}
    \caption{\textbf{An overview of CellMSA architecture.} For each target cell, we first use the CellMSA-Module to process the target cell together with its cellular context from different batches and biologically related cell types, summarizing cross-cell patterns into a context-dependent gene-pair representation. Then, in the GenePairformer module, the gene-pair representation is incorporated into the encoding of the target cell as an attention bias. Pretraining is
    conducted through joint optimization of three loss functions: masked gene expression modeling ($\mathcal{L}_\mathrm{MGM}$),  expression reconstruction ($\mathcal{L}_\mathrm{Rec}$) and cell contrastive learning ($\mathcal{L}_\mathrm{CCE}$).}
    \label{fig2}
\end{figure}

In this section, we describe the workflow of CellMSA, as illustrated in Figure~\ref{fig2}. We first curate a large-scale single-cell dataset and construct a hierarchically retrieved neighbor set for each target cell to form its corresponding \emph{MSA-like context}, as described in Section~\ref{3.1}. We then jointly model each target cell together with its context using the CellMSA-Module,
which extracts stable and differential patterns from the retrieved cells and compresses them into a gene-pair representation. 
Subsequently, in the GenePairformer module, this representation is incorporated into the target cell encoding as an attention bias, as described in Section~\ref{3.2}. Finally, Section~\ref{3.3} presents the pretraining objectives.

\subsection{Construction of MSA-like context for single cells}
\label{3.1}

Let  $X \in \mathbb{R}^{N \times G}$ denote a single-cell expression matrix, where $N$ denotes the number of cells and $G$ denotes the number of genes. 
CellMSA retrieves a set of similar cells with complementary information for each target cell and organizes them into a context analogous to a protein multiple sequence alignment (MSA).

To provide a diverse, biologically informative, and batch-robust context for the target cell, we sample context cells from the following three groups for target cell $c_i$:
$
\mathcal{N}_i =
\mathcal{N}^{\mathrm{same~batch}}_i
\cup
\mathcal{N}^{\mathrm{cross~batch}}_i
\cup
\mathcal{N}^{\mathrm{cross~type}}_i,
$
where $\mathcal{N}^{\mathrm{same~batch}}_i$ contains cells of the \textbf{same cell type and from the same batch} as the target cell, and is used to capture local similarity and support denoising;
$\mathcal{N}^{\mathrm{cross~batch}}_i$ contains cells of the \textbf{same cell type but from different batches}, and is used to encourage the model to extract biologically meaningful patterns that are stable across batches;
 $\mathcal{N}^{\mathrm{cross~type}}_i$ contains cells from \textbf{different but biologically similar cell types}, used to provide a contrastive background that highlights the distinctive properties of the target cell and further facilitates the learning of gene expression patterns associated with functional divergence.

To enable the model to distinguish context information from different sources, we assign each neighbor cell $c_j \in \mathcal{N}_i$ a relation-type label
$
r_{ij} \in
\{
\mathrm{same~batch},
\mathrm{cross~batch},
\mathrm{cross~type}
\}.
$
This label is encoded as a learnable relation embedding and added to the input representation of the corresponding neighbor cell.

Following the setup of scGPT \cite{cui2024scgpt}, each cell is represented as a sequence of gene expressions. The genes are selected from a predefined gene vocabulary and truncated or padded to a fixed length. Expression values are discretized into several bins: zero expression is preserved as a separate category, while nonzero expression values are partitioned into multiple levels according to within-cell quantiles. The discretized expression value, together with the gene identity, forms the model input, and a \texttt{<cls>} token is prepended to aggregate a cell-level representation.

\subsection{Model architecture}
\label{3.2}

CellMSA consists of three main components: an input embedding layer, a CellMSA-Module for extracting context-dependent gene-pair information from retrieved cells, and a GenePairformer as a target-cell encoder that incorporates the resulting pair representation.

\paragraph{Input embedding.}
For simplicity, in what follows we discuss only a single input instance. We organize the target cell $c_i$ together with its retrieved neighbors into an MSA-like input with $S = |\mathcal{N}_i| + 1$ rows, where the first row corresponds to the target cell. For the $s$-th row, let $x_{sg}$ denote the discretized expression value of gene $g$ after truncation and binning, and $r_s$ denote its relation type to the target cell. Let $\mathbf{e}_{sg} \in \mathbb{R}^{d}$ denote the initial representation of gene $g$ in the $s$-th cell. We define
\[
\mathbf{e}_{sg}
=
\begin{cases}
E_{\mathrm{gene}}(g) + E_{\mathrm{value}}(x_{sg}), & s = 1,\\
E_{\mathrm{gene}}(g) + E_{\mathrm{value}}(x_{sg}) + E_{\mathrm{rel}}(r_s), & s > 1.
\end{cases}
\]

Here, $E_{\mathrm{gene}}$ and $E_{\mathrm{rel}}$ are learnable gene-identity and relation-type embeddings, respectively. $E_{\mathrm{value}}$ is a learnable expression-value embedding function that maps the discretized expression bin $x_{sg}$ to a $d$-dimensional vector.

We also initialize the gene-pair representation $\mathbf{P}^{0} \in \mathbb{R}^{G \times G \times d_p}$, where $d_p$ is the pair representation dimension. Let $f_{\mathrm{pair}}$ be a learnable mapping. For genes $u$ and $v$, the initial pair representation is
$
\mathbf{p}_{uv}^{0}
=
f_{\mathrm{pair}}
\big(
E_{\mathrm{gene}}(u),\, E_{\mathrm{gene}}(v)
\big).
$

\paragraph{CellMSA-Module.}
The CellMSA-Module extracts cross-cell consistency and variation patterns from the MSA-like context and compresses them into the gene-pair representation. To reduce computation, we first project the input gene representations $\mathbf{e}_{sg}$ into a lower-dimensional  space, obtaining $\mathbf{m}_{sg}^{0} \in \mathbb{R}^{d'}$, where $d'<d$. $d'$ serves as the hidden dimension of the CellMSA-Module. This design allows the model to extract pairwise gene dependencies across multiple cells at relatively low cost before performing higher-dimensional modeling of the target cell. 
Let $\mathbf{m}^{l}$ and $\mathbf{P}^{l}$ denote the context representation and the gene-pair representation at layer $l$, respectively. In each layer, the module alternates between updating $\mathbf{P}^{l}$ and $\mathbf{m}^{l}$.

\emph{Update gene-pair representation $\mathbf{P}^{l}$.}
For genes $u$ and $v$, we aggregate their joint variation across context rows using an Outer-product Mean module \cite{abramson2024accurate}:
\[
\mathbf{p}_{uv}^{\,l+1}
=
\mathbf{p}_{uv}^{\,l}
+
W_{p}^{l}
\left(
\frac{1}{S}
\sum_{s=1}^{S}
\mathbf{a}_{su}^{\,l}
\otimes
\mathbf{b}_{sv}^{\,l}
\right),
\]
where $\mathbf{a}_{su}^{\,l}$ and $\mathbf{b}_{sv}^{\,l}$ are linear projections of $\mathbf{m}_{su}^{\,l}$ and $\mathbf{m}_{sv}^{\,l}$, respectively,  $\otimes$ denotes the outer product, and $W_{p}^{l}$ is a learnable mapping. 

\emph{Update context gene representation $\mathbf{m}^{l}$.}
Then, in the Pair-weighted Averaging module \cite{abramson2024accurate}, the updated pair representation is used as attention matrices to modulate information aggregation across gene positions:
\[
\mathbf{m}_{su}^{\,l+1}
=
\mathbf{m}_{su}^{\,l} +
W_{o}^{l}\,
\mathrm{Concat}_{h}
\left(
\gamma^{l,h}_{su}
\odot \sum_{v}\mathrm{softmax}_{v}\!\left(W_b^{\,l,h}\mathbf{p}_{uv}^{\,l+1}\right)\mathbf{v}_{sv}^{\,l,h}
\right),
\]
where $h$ indexes attention heads of CellMSA-Module, 
$\gamma^{l,h}_{su}$ denotes the gating representation obtained by applying a linear projection and a sigmoid transformation to $\mathbf{m}_{su}^{\,l}$, 
$\mathbf{v}_{sv}^{\,l,h}$ denotes the value representation obtained from $\mathbf{m}_{sv}^{\,l}$, $W_{o}^{l}$ and $W_b^{\,l,h}$ are learnable mappings. The structures of the Outer-product Mean module and Pair-weighted Averaging module are detailed in Appendix \ref{app:architecture}.

\paragraph{GenePairformer.}
After obtaining the final gene-pair representation $\mathbf{P}^{L_{\mathrm{MSA}}}$ from the CellMSA-Module, we project it into head-specific pair representations, denoted by $\mathbf{R}^{0}$, which serve as the pair input to the GenePairformer. Meanwhile, we use the gene embeddings of the target cell, $\mathbf{e}_{1g}$, as the initial sequence representation $\mathbf{h}^{0}$.

At each layer $L$, the GenePairformer jointly updates the pair representation $\mathbf{R}^{L}$ and the sequence representation $\mathbf{h}^{L}$. We adopt a simple yet effective pair-aware design \cite{unimol, esmaa, zhao2025stofm} in which the pair representation is injected into self-attention as an attention bias, allowing the target-cell encoder to explicitly leverage the gene-gene dependencies extracted from the multi-cell context. The resulting attention logits are then used to further refine the pair representation, enabling iterative interaction between sequence features and pairwise structure. Specifically, the following modifications are made based on the standard Transformer \cite{vaswani2017attention}:
\[
\mathbf{R}_{uv}^{L+1,H} =
\mathbf{R}_{uv}^{L,H}
+
\frac{\mathbf{Q}_{u}^{L,H}\left(\mathbf{K}_{v}^{L,H}\right)^\top}{\sqrt{d_H}},
\]
\[
\mathrm{Attention}_{uv}^{L,H} =
\mathrm{softmax}_{v}\!\left(\mathbf{R}_{uv}^{L+1,H}\right)\mathbf{V}_{v}^{L,H},
\]
where $H$ indexes the attention head, and $d_H$ denotes the hidden dimension of each head. $\mathbf{Q}^{L,H}$, $\mathbf{K}^{L,H}$, and $\mathbf{V}^{L,H}$ denote the query, key, and value representations of head $H$, obtained by linear projections of $\mathbf{h}^{L}$, respectively. Finally, we take the output corresponding to the \texttt{<cls>} token and project it into the final cell representation $\mathbf{z}_i \in \mathbb{R}^{d}$.
In this way, the model performs fine-grained modeling of the target cell while explicitly leveraging gene relationship information extracted from the MSA-like context.

\subsection{Pretraining objectives}
\label{3.3}

CellMSA is jointly pretrained with three objectives tailored to single-cell data: \textbf{masked gene expression modeling}, \textbf{cell expression profile reconstruction}, and \textbf{cell-level contrastive learning}. These objectives encourage the model to recover locally missing expression, capture global transcriptomic structure, and learn biologically discriminative cell representations, respectively.

\paragraph{Masked gene expression modeling.}
We randomly mask a subset of gene positions and use the model outputs at these positions to predict their original expression bins. To prevent trivial copying from neighboring cells, the same gene positions are masked simultaneously in both the target cell and its contextual rows. Let $\Omega$ denote the set of masked gene positions. The cross-entropy loss is:
\[
\mathcal{L}_{\mathrm{MGM}}
=
\frac{1}{|\Omega|}\sum_{g\in\Omega}
\mathrm{CrossEntropy}(x_{i,g}, \hat{x}_{i,g})
\]

\paragraph{Cell expression profile reconstruction.}
To encourage the cell representation to preserve global transcriptional information, we reconstruct gene expression from the final cell embedding $\mathbf{z}_i$. For gene $g$ from a randomly selected subset $\Gamma$, we concatenate $\mathbf{z}_i$ with the initial gene embedding $E_{\mathrm{gene}}(g)$, feed the result into a reconstruction head $f_{\mathrm{rec}}$, and compute the cross-entropy with the ground-truth binned expression:
\[
\mathcal{L}_{\mathrm{Rec}}
=
\frac{1}{|\Gamma|}\sum_{g\in\Gamma}
\mathrm{CrossEntropy}(x_{i,g}, f_{\mathrm{rec}}
\big(
\mathbf{z}_i, E_{\mathrm{gene}}(g)
\big))
\]

\paragraph{Cell-level contrastive learning.}
We apply a contrastive objective to improve the biological discriminability of cell representations. For each target cell, we sample a positive example from the same cell type and encourage the two representations to be close while separating them from other candidate cells from the same batch. 
Let $\mathcal{B}$ denote the candidate set in the current contrastive batch, $\tau$ denote the temperature parameter, and $\mathrm{sim}$ denote cosine similarity. The InfoNCE loss \cite{oord2018representation, MOCO_He2019} is:
\[
\mathcal{L}_{\mathrm{CCE}}
=
-\log
\frac{
\exp(\mathrm{sim}(\mathbf{z}_i,\mathbf{z}_i^{+})/\tau)
}{
\sum_{j \in \mathcal{B}}
\exp(\mathrm{sim}(\mathbf{z}_i,\mathbf{z}_j)/\tau)
},
\]

The overall pretraining objective is
$
\mathcal{L}
=
\mathcal{L}_{\mathrm{MGM}}
+
\lambda_{1}\mathcal{L}_{\mathrm{Rec}}
+
\lambda_{2}\mathcal{L}_{\mathrm{CCE}},
$
where $\lambda_{1}$ and $\lambda_{2}$ control the weights of the reconstruction and contrastive losses, respectively. We pretrain CellMSA on approximately 109 million human cell observations, including 65.6 million primary observations, enabling it to learn transferable cell representations.

\section{Experiments}
\label{sec:experiments}

\subsection{Experiment settings}
\label{sec:experiment_settings}

\paragraph{Pretraining setup.}

We curated a pretraining corpus from the CELLxGENE database, comprising 109 million human cell observations from 2090 datasets and 818 cell types, including 65.6 million primary observations \cite{czi2025cz}. All datasets used for downstream evaluation are excluded from the pretraining corpus before model training. Primary observations are identified by the \texttt{is\_primary\_data} flag; the corpus is not restricted to these observations. Appendix~\ref{app:pretraining_setting} explains the retained non-primary observations and their context-conditioned use. CELLxGENE database contains both batch and cell type annotations, which we leverage to construct $\mathcal{N}^{\mathrm{same~batch}}$ and $\mathcal{N}^{\mathrm{cross~batch}}$. To systematically model cell-type relationships, we aggregate each cell type into a global representation via a cell-count-weighted average of expression profiles across all relevant batches. Based on a pairwise cosine similarity matrix of these representations, we apply hierarchical clustering with Ward's linkage \cite{murtagh2011ward} and rank members within each cluster by similarity to define structured neighborhoods. This hierarchy guides the sampling of $\mathcal{N}^{\mathrm{cross~type}}$ for MSA-like context construction.
Following the retrieval strategy in Section~\ref{3.1}, the MSA-like context for each target cell $i$ is formed by retrieving 16 $\mathcal{N}^{\mathrm{same~batch}}_i$, 16 $\mathcal{N}^{\mathrm{cross~batch}}_i$, and 8 $\mathcal{N}^{\mathrm{cross~type}}_i$. We pretrain the model for one epoch using 4 NVIDIA A800 GPUs, which takes approximately 20 days. Detailed pretraining hyperparameters are provided in Appendix~\ref{app:pretraining_setting}.

\paragraph{Baseline selection and downstream task setup.}

For baseline comparisons, we benchmark against a diverse suite of representative models, broadly categorized into three distinct groups. The first encompasses traditional statistical methods, represented by PC-HVG \cite{butler2018integrating}. The second comprises deep generative models, represented by scVI \cite{lopez2018deep}. The final group consists of recent large-scale single-cell foundation models, namely scGPT \cite{cui2024scgpt}, Geneformer \cite{geneformer}, STATE-SE \cite{adduri2025predicting}, and Stack \cite{dong2026stack}. We additionally evaluate CellPLM \cite{wen2023cellplm}, using its official implementation and pretrained checkpoint, on cell type and PT cell state classification.

\subsection{Single-cell batch integration}
\label{sec:batch_integration}

In the single-cell batch integration task, the primary goal is to remove technical variation across batches, platforms, and donors while strictly preserving genuine biological structure, particularly cell-type identity and local neighborhood organization. To evaluate this, we utilized the Tabula Sapiens dataset \cite{Tabula_Sapiens}, a comprehensive human single-cell transcriptomic atlas. This dataset comprises 483,152 cells spanning 24 distinct tissues and organs from 15 human donors, including 475 cell types.
We employed the scib-metrics framework \cite{scib} to quantify integration quality across two dimensions: batch correction and biological conservation. Specifically, batch correction is assessed using BRAS, iLISI, kBET, graph connectivity (Conn), and principal component regression comparison (PCR); biological conservation is evaluated via NMI, ARI, ASW, and cLISI. For detailed definitions of these metrics, please refer to Appendix \ref{app_metric_batch}.

The main integration benchmark uses \emph{label-informed} context retrieval for CellMSA and Stack, corresponding to atlas alignment with available cell-type annotations. The remaining baselines do not use these labels for representation extraction, so comparisons across these settings involve different information access. Appendix~\ref{app:label_free_integration} additionally reports a label-free comparison on Bladder, in which context retrieval uses expression-based neighbors without cell-type labels.

\begin{figure}[t]
    \centering
    \includegraphics[width=1\linewidth]{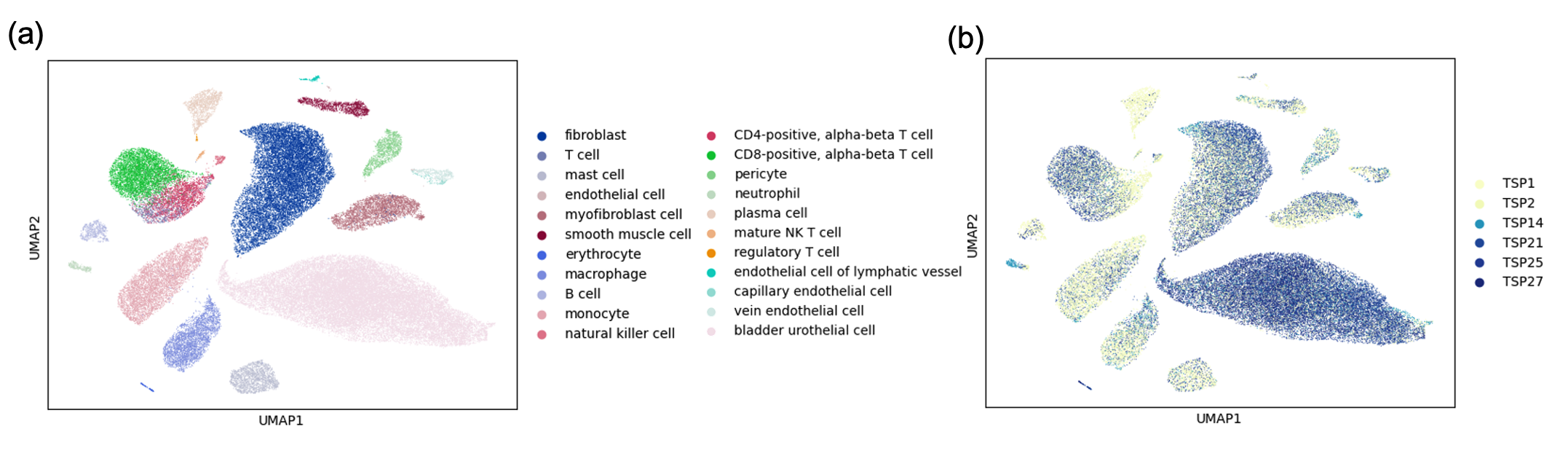}
    \caption{UMAP visualization of the CellMSA latent space on the Tabula Sapiens-Bladder dataset using label-informed context retrieval, colored by (a) cell types and (b) donor ID.}
    \label{fig:umap}
\end{figure}

As indicated by the tissue-averaged results in Table \ref{tab:baseline_comparison}, CellMSA achieves superior batch mixing across technical origins while preserving biological states, with the highest aggregate score among the evaluated methods under the stated information-access settings (see Appendix \ref{app:batch_expr} for detailed tissue-specific results). Consequently, these representations provide a strong foundation for diverse downstream tasks. Figure \ref{fig:umap} presents a UMAP visualization of the CellMSA latent space to intuitively demonstrate its batch integration performance using the Tabula Sapiens-Bladder dataset.

\begin{table}[t]
    \centering
    \caption{Results of single-cell batch integration. CellMSA and Stack use label-informed context retrieval; the other methods do not use cell-type labels for representation extraction. A label-free comparison is reported in Appendix~\ref{app:label_free_integration}.}
    \label{tab:baseline_comparison}
    \resizebox{\textwidth}{!}{
    \begin{tabular}{lcccccccccccc}
        \toprule
         & \multicolumn{4}{c}{Biological conservation} & \multicolumn{5}{c}{Batch correction} & \multicolumn{3}{c}{\textbf{Total score}} \\
        \cmidrule(r){2-5} \cmidrule(r){6-10} \cmidrule{11-13}
        \textbf{Models} & NMI & ARI & ASW & cLISI & BRAS & iLISI & kBET & Conn & PCR & \textbf{Batch} & \textbf{Bio} & \textbf{Total} \\
        \midrule

        PC-HVG     & 0.731 & 0.594 & 0.574 & \textbf{1.000} & 0.621 & 0.028 & 0.366 & 0.780 & 0.000 & 0.359 & 0.725 & 0.578 \\
        scVI       & 0.741 & 0.592 & 0.562 & \textbf{1.000} & 0.649 & 0.046 & 0.363 & 0.822 & 0.190 & 0.414 & 0.724 & 0.600 \\
        scGPT      & 0.730 & 0.558 & 0.569 & \textbf{1.000} & 0.670 & 0.033 & 0.346 & 0.789 & 0.128 & 0.393 & 0.714 & 0.586 \\
        Geneformer & 0.709 & 0.539 & 0.531 & 0.999 & 0.751 & 0.035 & 0.344 & 0.739 & 0.222 & 0.418 & 0.694 & 0.584 \\
        STATE-SE   & 0.730 & 0.553 & 0.530 & \textbf{1.000} & \textbf{0.878} & 0.010 & 0.315 & 0.503 & 0.417 & 0.425 & 0.703 & 0.592 \\
        Stack      & 0.772 & 0.678 & 0.513 & 0.999 & 0.865 & 0.088 & 0.377 & 0.831 & \textbf{0.573} & 0.547 & 0.740 & 0.663 \\
        \textbf{CellMSA} & \textbf{0.895} & \textbf{0.805} & \textbf{0.699} & \textbf{1.000} & 0.740 & \textbf{0.242} & \textbf{0.627} & \textbf{0.945} & 0.278 & \textbf{0.566} & \textbf{0.850} & \textbf{0.736} \\
        \bottomrule
    \end{tabular}
    }
\end{table}

\subsection{Cell type and cell state classification }
\label{sec:cell_classification}

To further assess whether CellMSA can capture biologically meaningful fine-grained cellular variation, we evaluated the model on two classification tasks at different levels of granularity: cell type annotation on Tabula Sapiens-Blood and proximal tubule (PT) cell state classification on Kidney Atlas~\cite{de2021rationale}. The first task evaluates the transferability of learned representations to broad immune cell identities, whereas the second task focuses on disease-associated transcriptional states within a single kidney epithelial lineage. 
To avoid label leakage, we construct the MSA context for validation and test cells using HVG + PCA + KNN \cite{cover1967nearest}. For both tasks, we train a lightweight MLP classification head on top of the cell embeddings. 
We randomly split the data into training/validation/test sets by donor ID and conduct five random experiments. Table \ref{tab:performance_metrics} reports the corresponding means and standard deviations. See Appendix~\ref{app:downstream_dataset} and \ref{app:downstream_setting} for more details.

\paragraph{Cell type annotation on Tabula Sapiens-Blood.}
We consider a standard cell type annotation task using labels from Tabula Sapiens-Blood. This dataset contains multiple closely related cell subtypes, providing a challenging benchmark. Compared with conventional single-cell foundation model baselines, CellMSA achieves the best performance, indicating its ability to learn cell-type-specific gene expression patterns from cellular context. The additional CellPLM baseline reaches 0.906 macro-F1, compared with 0.912 for CellMSA. Appendix~\ref{app:gene_dropout} further evaluates embedding stability under random removal of nonzero genes.

\paragraph{PT cell state classification on Kidney Atlas.}
We evaluate a disease-focused classification task on proximal tubule (PT) cells from Kidney Atlas. Under disease conditions, PT cells can be divided into three cell states: aPT, dPT, and dPT/DTL, corresponding to normal state, injured state, and severely injured state with transcriptional identity drift, respectively. 
This task requires the models to distinguish disease-related cell states within the same cell type. In this setting, CellMSA achieves the best performance, suggesting that cellular context helps capture subtle disease-associated state differences.

\begin{table}[t]
    \centering
    \caption{Results of cell type and cell state classification. Entries with $\pm$ report mean and standard deviation over five runs.}
    \label{tab:performance_metrics}
    \resizebox{\textwidth}{!}{
    \begin{tabular}{lcccccc}
        \toprule
         & \multicolumn{3}{c}{Cell type annotation} &  \multicolumn{3}{c}{PT cell state classification} \\
        \cmidrule(r){2-4} \cmidrule(r){5-7}
        \textbf{Model} & \textbf{Accuracy} & \textbf{Macro F1} & \textbf{Weighted F1} & \textbf{Accuracy} & \textbf{Macro F1} & \textbf{Weighted F1} \\
        \midrule
        scVI        & $0.905 \pm 0.002$ & $0.809 \pm 0.006$ & $0.908 \pm 0.002$ & $0.901 \pm 0.011$ & $0.876 \pm 0.014$ & $0.900 \pm 0.012$ \\
        scGPT       & $0.895 \pm 0.004$ & $0.784 \pm 0.008$ & $0.898 \pm 0.003$ & $0.926 \pm 0.007$ & $0.895 \pm 0.012$ & $0.927 \pm 0.007$ \\
        Geneformer  & $0.896 \pm 0.006$ & $0.803 \pm 0.012$ & $0.899 \pm 0.004$ & $0.930 \pm 0.008$ & $0.903 \pm 0.016$ & $0.930 \pm 0.008$ \\
        STATE-SE    & $0.908 \pm 0.009$ & $0.815 \pm 0.017$ & $0.910 \pm 0.007$ & $0.928 \pm 0.009$ & $0.912 \pm 0.009$ & $0.928 \pm 0.010$ \\
        Stack       & $0.914 \pm 0.003$ & $0.822 \pm 0.008$ & $0.916 \pm 0.002$ & $0.936 \pm 0.008$ & $0.909 \pm 0.009$ & $0.938 \pm 0.008$ \\
        CellPLM     & $0.931 \pm 0.003$ & $0.906 \pm 0.010$ & $0.932 \pm 0.003$ & $0.850 \pm 0.010$ & $0.794 \pm 0.012$ & $0.852 \pm 0.010$ \\
        \textbf{CellMSA}  & $\mathbf{0.962 \pm 0.002}$ & $\mathbf{0.912 \pm 0.005}$ & $\mathbf{0.962 \pm 0.002}$ & $\mathbf{0.958 \pm 0.004}$ & $\mathbf{0.931 \pm 0.006}$ & $\mathbf{0.958 \pm 0.004}$ \\
        \bottomrule
    \end{tabular}
    }
\end{table}

\subsection{Perturbation prediction}
\label{sec:perturbation_prediction}
Predicting cellular responses to perturbations is essential for understanding complex regulatory mechanisms and accelerating therapeutic discovery \cite{wei2026benchmarking}. This task involves modeling transcriptomic shifts that occur when cells are subjected to external stimuli such as chemical compounds or genetic modifications. 
Recently, STATE \cite{adduri2025predicting} establishes a robust framework for evaluating the performance of single-cell foundation models on this task. This framework allows any model to be combined with a State Transition (STATE-ST) module for perturbation prediction. Following this setup, we evaluate CellMSA and compare it with the original gene expression profiles, STATE-SE, and Stack.

We use a filtered subset of the Replogle dataset \cite{replogle2022mapping}, which includes four human cell lines and 100 gene perturbations (see Appendix \ref{app:downstream_dataset} and \ref{app:downstream_setting} for details).
We use Pearson $\Delta$ to directly assess the consistency between the predicted and actual perturbation effects, and employ Spearman-FC, PRAUC, and DE Overlap to evaluate predictive accuracy with respect to differentially expressed (DE) genes. These evaluation metrics are adopted from Cell-Eval \cite{adduri2025predicting}.
As shown in Table \ref{tab:perturbation_results}, compared to the baseline models, CellMSA exhibits higher agreement between predicted and real perturbation effects, together with higher fidelity in predicting DE genes. This confirms that CellMSA provides a more robust and biologically informative latent space for modeling complex cellular transitions.

\begin{table}[t]
    \centering
    \caption{Results of perturbation prediction.}
    \label{tab:perturbation_results}
    \resizebox{0.8\textwidth}{!}{
    \begin{tabular}{lcccc}
        \toprule
        \textbf{Model} & \textbf{Pearson $\Delta$ } & \textbf{PRAUC} & \textbf{Spearman-FC} & \textbf{DE Overlap} \\
        \midrule
        Expression + STATE-ST   & 0.398 & 0.276 & 0.404 & 0.139 \\
        STATE-SE + STATE-ST     & 0.353 & 0.287 & 0.365 & 0.180 \\
        Stack + STATE-ST        & 0.358 & 0.293 & 0.372 & 0.148 \\
        \textbf{CellMSA + STATE-ST} & \textbf{0.433} & \textbf{0.334} & \textbf{0.431} & \textbf{0.215} \\
        \bottomrule
    \end{tabular}
    }
    \vspace{1mm}
\end{table}

\subsection{Interpretability of gene-pair representations}

\begin{figure}
    \centering
    \includegraphics[width=1\linewidth]{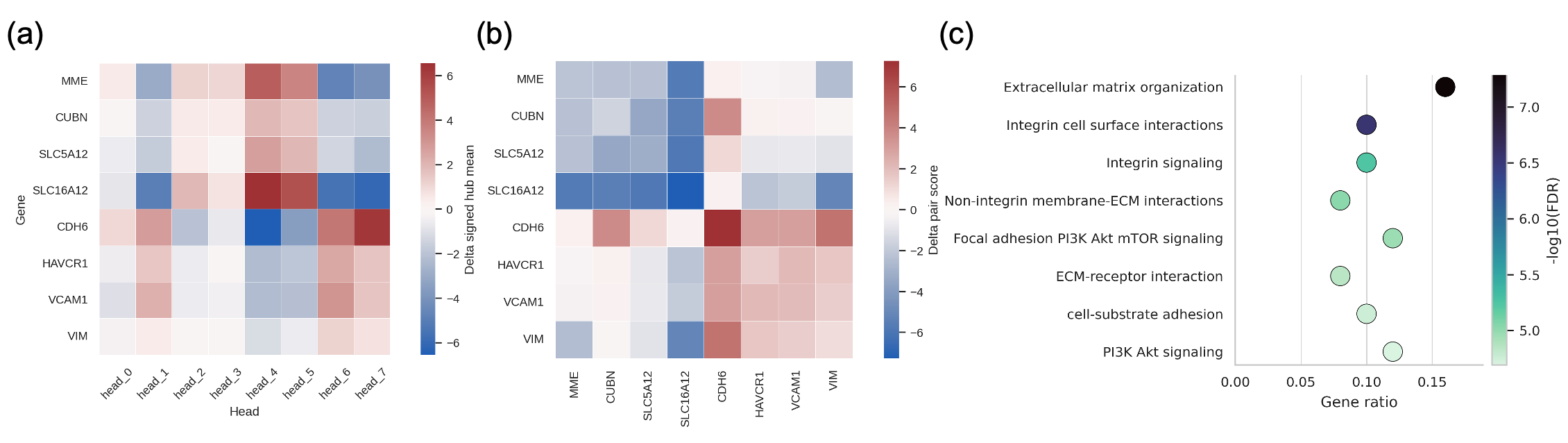}
    \caption{\textbf{Disease-associated gene network captured by gene-pair representations.} (a) Heatmap of differential hub scores for PT marker genes. (b) Gene-pair heatmap of PT marker genes within head 7. (c) Gene Ontology (GO) enrichment analysis of the top-ranked genes derived from head 7.}
    \label{fig4}
\end{figure}

The gene-pair representations derived from the CellMSA-Module enable us to capture state-dependent gene-gene relationships at the single-cell level. By aggregating these representations across cells within the same biological condition, CellMSA extracts cell-type- and state-specific gene-pair activation patterns (see Appendix~\ref{app:interpretability}). 
Next, we investigate whether these pairwise patterns capture disease-associated network reconfiguration. We analyze PT cells from the Kidney Atlas under disease versus healthy conditions. For each head, we compute the differential gene-pair representations between disease and healthy states. We evaluate their alignment with established marker genes for renal health (MME, CUBN, SLC5A12, SLC16A12) and disease (CDH6, HAVCR1, VCAM1, VIM) \cite{christensen2013bowel, gerhardt2021single, verouti2021solute, lee2015deep, moreno2021post}. The results demonstrate that heads 6 and 7 preferentially capture transcriptional dependencies associated with the disease state, while heads 4 and 5 are more closely aligned with homeostatic signatures (Figure~\ref{fig4}a). The gene-pair heatmap of PT marker genes within head 7 explicitly displays their altered interaction patterns (Figure~\ref{fig4}b).

Focusing on the disease-associated modes (head 7), we identify a subset of genes that exhibit the most significant shifts in pairwise connectivity between healthy and diseased conditions. Notably, a substantial proportion of these high-connectivity genes overlap with known proximal tubule injury markers, suggesting that CellMSA's representations are anchored in robust pathological signals. To further resolve the biological processes captured by these heads, we conducted Gene Ontology (GO) \cite{ashburner2000gene} enrichment analysis on the top-ranked genes within head 7. The results reveal a significant enrichment of well-established disease-associated pathways, including extracellular matrix (ECM) reorganization, altered cell-substrate interactions (e.g., integrin signaling), and cellular dedifferentiation (e.g., cell and tube morphogenesis) \cite{rayego2021interplay, elwakiel2024factor, romero2025dynamics} (Figure~\ref{fig4}c). This functional profile suggests a transition of proximal tubule cells toward a dedifferentiated, profibrotic state, reflecting a maladaptive repair process in response to kidney injury. Overall, these findings demonstrate that CellMSA can capture disease-driven network variations as biologically interpretable latent modes. An external comparison with STRING functional associations is provided in Appendix~\ref{app:string_validation}.

\subsection{Ablation study}

\begin{figure}[t]
\centering
\begin{minipage}{0.55\textwidth}
\centering
\captionof{table}{Results of ablation studies.}
\label{tab:ablation}
\resizebox{\linewidth}{!}{
\begin{tabular}{lcccc}
\toprule
& \multicolumn{2}{c}{PT cell state classification} &  \multicolumn{2}{c}{Perturbation prediction} \\
\cmidrule(r){2-3} \cmidrule(r){4-5}
Method & Accuracy & Macro F1 & Pearson $\Delta$ & Spearman-FC \\
\midrule
w/o CellMSA-Module             & 0.854 & 0.811 & 0.355 & 0.366 \\
w/o Context             & 0.878 & 0.817 & 0.389 & 0.367 \\
Only Same-batch Context & 0.929 & 0.893 & 0.400 & 0.397 \\
\textbf{Full Model}     & \textbf{0.958} & \textbf{0.931} & \textbf{0.433} & \textbf{0.431} \\
\bottomrule
\end{tabular}}
\end{minipage}
\hfill
\begin{minipage}{0.44\textwidth}
\centering
\includegraphics[width=\linewidth]{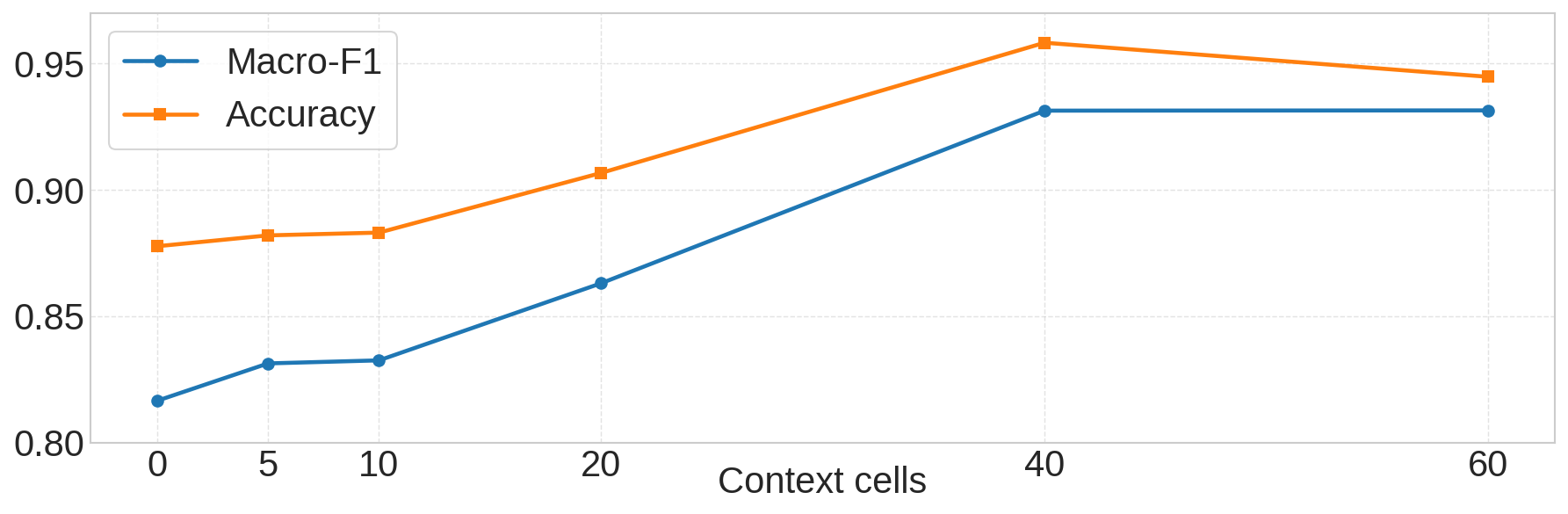}
\caption{Ablation study of the context size.}
\label{fig:curve}
\end{minipage}
\end{figure}

We conduct ablation studies to verify that the performance gains of CellMSA come from effectively leveraging the MSA context. First, we completely remove the CellMSA-Module and the pair representation, in which case the GenePairformer degenerates into a standard Transformer. Second, while keeping the architecture unchanged, we remove the contextual cells from the input. Third, following prior work, we use only the same-batch context, in order to evaluate the benefit of using more informative contexts. These ablation studies are conducted on PT cell state classification and perturbation prediction.
As shown in Table ~\ref{tab:ablation}, the full model consistently achieves the best performance across all tasks, and the ablation results demonstrate the performance gains resulting from incorporating the information-rich MSA context.

We further study the effect of the number of context cells on PT cell state classification task. As the context size increases from zero, the downstream performance first improves and then gradually saturates. When more than 40 context cells are used, further increasing the context size no longer brings substantial performance gains (as shown in Figure \ref{fig:curve}).
In addition, as part of hyperparameter analysis, we also study the effects of the pretraining objectives; details are provided in Appendix \ref{app:ablation_objectives}.

\section{Conclusions and limitations}
\label{sec:conclusion}

We presented CellMSA, an MSA-inspired framework for single-cell representation learning. By incorporating informative contextual cells and summarizing cross-cell patterns into gene-pair representations, CellMSA enables more robust and fine-grained single-cell modeling. It achieves strong performance across various downstream tasks, opening up new perspectives for single-cell data analysis.
Currently, CellMSA still has some limitations.
While CellMSA captures informative gene expression patterns by comparing consistency and variation across cellular contexts, these patterns mainly reflect statistical associations rather than explicit causal regulatory relationships. Our future work may incorporate prior knowledge of gene regulation and causal machine learning methods to further strengthen the biological grounding and interpretability.

\begin{ack}
This research is supported by the Innovative Drug Research and Development National Science and Technology Major Project (No. 2025ZD1803101) and PharMolix Inc.
\end{ack}

\bibliographystyle{unsrtnat}
\bibliography{cellmsa}

\newpage
\appendix
\onecolumn
\setcounter{section}{0}
\setcounter{equation}{0}
\setcounter{subsection}{0}
\setcounter{table}{0}
\setcounter{figure}{0}
\renewcommand{\theequation}{A.\arabic{equation}}
\renewcommand{\thetable}{A.\arabic{table}}
\renewcommand{\thefigure}{A.\arabic{figure}}

\section*{Appendix}

\section{Details and analysis of model components}

\subsection{Details of Outer-product Mean and Pair-weighted Averaging}
\label{app:architecture}

\begin{figure}[ht]
    \centering
    \includegraphics[width=1\linewidth]{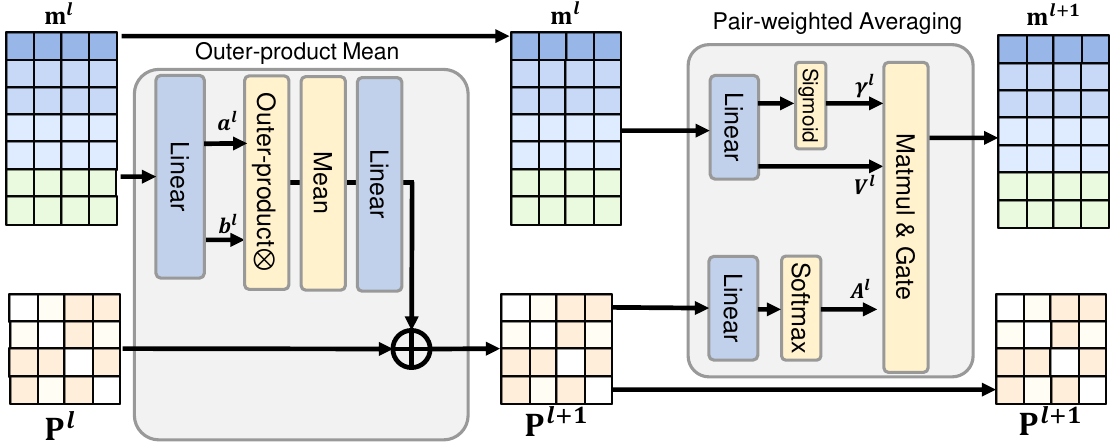}
    \caption{\textbf{The structure of Outer-product Mean and Pair-weighted Averaging in CellMSA-Module.}}
    \label{figa1}
\end{figure}

Figure~\ref{figa1} provides detailed illustrations of the internal structures of the Outer-product Mean module and Pair-weighted Averaging module shown in Figure~\ref{fig2}.

\paragraph{Outer-product Mean}
The outer-product mean operation updates the gene-pair representation by aggregating cross-cell co-variation patterns from the MSA-like context. For each gene pair $(u,v)$, we project the context representation $m^l_{su}$ into two low-dimensional vectors $a^l_{su}$ and $b^l_{sv}$, and average their outer products over all context rows:\[o^l_{uv}=\frac{1}{S}\sum_{s=1}^{S} a^l_{su}\otimes b^l_{sv}.\]The result is then projected back to the pair space and added to the previous pair representation:\[p^{l+1}_{uv}=p^l_{uv}+W^l_p \operatorname{Flatten}(o^l_{uv}).\]This operation enables CellMSA to summarize context-dependent gene-gene dependencies from related cells while keeping the computation in a reduced hidden dimension.

\paragraph{Pair-weighted Averaging}
Pair-weighted averaging feeds the updated gene-pair representation back into the context gene representations. For each attention head, the pair representation is projected into attention logits:
\[A^{l,h}_{uv}=\operatorname{softmax}_v\left(W^{l,h}_b p^{l+1}_{uv}\right).\]

Each gene representation is then updated by aggregating value vectors from other genes using these pair-induced weights:
\[
\mathbf{m}^{l+1}_{su}=\mathbf{m}^l_{su}+
W^l_o\operatorname{Concat}_h\left(\gamma^{l,h}_{su}
\odot \sum_v A^{l,h}_{uv} \mathbf{v}^{l,h}_{sv}\right).
\]

This step allows the CellMSA-Module to iteratively exchange information between gene-level context representations and context-dependent gene-pair representations.

\subsection{Complexity analysis}
\label{app:complexity}

We analyze the complexity of the key modules in CellMSA. Let \(S=|\mathcal{N}_i|+1\) denote the number of rows in the MSA-like input, including the target cell and its contextual cells, and let \(G\) denote the number of selected genes. The CellMSA-Module operates on the low-dimensional context representation \(m^l_{sg}\in\mathbb{R}^{d'}\), where \(d'<d\), and maintains a gene-pair representation \(P^l\in\mathbb{R}^{G\times G\times d_p}\). The GenePairformer encodes only the target cell with hidden states \(h^L\in\mathbb{R}^{G\times d}\).

\paragraph{Outer-product mean.} The outer-product mean operation updates the pair representation \(P^l\) by aggregating cross-cell statistics for each gene pair \((u,v)\). Since this operation is computed for all \(G^2\) gene pairs and aggregates over \(S\) context rows, its dominant cost is quadratic in the number of genes and linear in the number of context cells. In our implementation of CellMSA, we set the projection dimension of \(a^l_{su}\) and \(b^l_{sv}\) to \(d_p\). Therefore, the outer-product aggregation has time complexity \(O(SG^2d_p^2)\). 

\paragraph{Pair-weighted averaging.} The pair-weighted averaging operation feeds the updated pair representation back to the context gene representations. This operation first projects \(P^l\) into gene-gene attention weights and then applies these weights to all \(S\) rows of context representations. Its dominant time complexity is \(O(G^2d_p+SG^2d')\).

\paragraph{GenePairformer.} GenePairformer encodes only the target cell, rather than all \(S\) context rows. Although it uses the pair representation \(R^L\) as an attention bias, this does not change the asymptotic complexity compared with a standard Transformer over the target-cell gene sequence. 
Therefore, each GenePairformer layer has the same asymptotic time complexity as a standard Transformer layer, namely \(O(Gd^2+G^2d)\). 

\paragraph{Overall complexity.} Overall, the main additional cost of CellMSA compared with a standard single-cell Transformer comes from the CellMSA-Module:
\[
O\left(L_{\mathrm{MSA}}\left(SG^2d_p^2+SG^2d'\right)\right).
\]
The GenePairformer keeps the same asymptotic complexity as a standard Transformer encoder applied to the target cell:
\[
O\left(L_{\mathrm{GPF}}(Gd^2+G^2d)\right).
\]
This design allows CellMSA to extract context-dependent gene-pair information from multiple cells in the reduced dimension \(d'\), while performing the more expensive high-dimensional Transformer encoding only on the target cell.

\section{More experimental results}

\subsection{Batch integration experiments}
\label{app:batch_expr}
We applied the scib-metrics benchmarking pipeline independently to each tissue in Tabula Sapiens dataset. This section presents the detailed, tissue-level evaluation results that constitute the macro-averaged scores reported in the main text.  In these tissue-level tables, CellMSA and Stack use label-informed context retrieval, whereas the other methods do not use cell-type labels for representation extraction. A separate label-free comparison follows below.

\begin{table}[htbp]
    \centering
    \caption{Performance comparison of different models for single-cell batch integration on the Bladder dataset.}
    \label{tab:benchmark_bladder}
    \resizebox{\textwidth}{!}{
    \begin{tabular}{lcccccccccccc}
        \toprule
         & \multicolumn{4}{c}{Biological conservation} & \multicolumn{5}{c}{Batch correction} & \multicolumn{3}{c}{\textbf{Total score}} \\
        \cmidrule(r){2-5} \cmidrule(r){6-10} \cmidrule{11-13}
        \textbf{Models} & NMI & ARI & ASW & cLISI & BRAS & iLISI & kBET & Conn & PCR & \textbf{Batch} & \textbf{Bio} & \textbf{Total} \\
        \midrule
        PC-HVG & 0.784 & 0.735 & 0.647 & \textbf{1.000} & 0.661 & 0.038 & 0.212 & 0.789 & 0.000 & 0.340 & 0.791 & 0.611 \\
        scVI & 0.812 & 0.748 & 0.601 & \textbf{1.000} & 0.702 & 0.097 & 0.237 & 0.866 & 0.429 & 0.466 & 0.790 & 0.661 \\
        scGPT & 0.744 & 0.539 & 0.599 & \textbf{1.000} & 0.711 & 0.045 & 0.195 & 0.817 & 0.298 & 0.413 & 0.720 & 0.598 \\
        Geneformer & 0.744 & 0.602 & 0.566 & \textbf{1.000} & 0.770 & 0.031 & 0.217 & 0.748 & 0.425 & 0.438 & 0.728 & 0.612 \\
        STATE-SE & 0.698 & 0.432 & 0.582 & \textbf{1.000} & \textbf{0.884} & 0.003 & 0.210 & 0.438 & 0.457 & 0.399 & 0.678 & 0.566 \\
        Stack & 0.833 & \textbf{0.868} & 0.532 & \textbf{1.000} & 0.873 & 0.089 & 0.260 & 0.816 & \textbf{0.627} & 0.533 & 0.808 & 0.698 \\
        \textbf{CellMSA} & \textbf{0.915} & 0.781 & \textbf{0.772} & \textbf{1.000} & 0.746 & \textbf{0.347} & \textbf{0.541} & \textbf{0.944} & 0.276 & \textbf{0.571} & \textbf{0.867} & \textbf{0.748} \\
        \bottomrule
    \end{tabular}
    }
\end{table}

\begin{table}[htbp]
    \centering
    \caption{Performance comparison of different models for single-cell batch integration on the Blood dataset.}
    \label{tab:benchmark_blood}
    \resizebox{\textwidth}{!}{
    \begin{tabular}{lcccccccccccc}
        \toprule
         & \multicolumn{4}{c}{Biological conservation} & \multicolumn{5}{c}{Batch correction} & \multicolumn{3}{c}{\textbf{Total score}} \\
        \cmidrule(r){2-5} \cmidrule(r){6-10} \cmidrule{11-13}
        \textbf{Models} & NMI & ARI & ASW & cLISI & BRAS & iLISI & kBET & Conn & PCR & \textbf{Batch} & \textbf{Bio} & \textbf{Total} \\
        \midrule
        PC-HVG & 0.714 & 0.485 & 0.627 & \textbf{1.000} & 0.614 & 0.021 & 0.332 & 0.808 & 0.000 & 0.355 & 0.706 & 0.566 \\
        scVI & 0.745 & 0.507 & 0.598 & \textbf{1.000} & 0.667 & 0.036 & 0.321 & 0.848 & 0.285 & 0.431 & 0.713 & 0.600 \\
        scGPT & 0.721 & 0.478 & 0.630 & \textbf{1.000} & 0.670 & 0.029 & 0.277 & 0.834 & 0.166 & 0.395 & 0.707 & 0.582 \\
        Geneformer & 0.724 & 0.633 & 0.581 & \textbf{1.000} & 0.753 & 0.032 & 0.255 & 0.713 & 0.303 & 0.411 & 0.734 & 0.605 \\
        STATE-SE & 0.741 & 0.541 & 0.544 & \textbf{1.000} & \textbf{0.858} & 0.000 & 0.230 & 0.360 & 0.502 & 0.390 & 0.707 & 0.580 \\
        Stack & 0.791 & 0.726 & 0.522 & \textbf{1.000} & 0.853 & 0.072 & 0.250 & 0.824 & \textbf{0.691} & \textbf{0.538} & 0.760 & 0.671 \\
        \textbf{CellMSA} & \textbf{0.897} & \textbf{0.869} & \textbf{0.702} & \textbf{1.000} & 0.758 & \textbf{0.180} & \textbf{0.660} & \textbf{0.927} & 0.162 & 0.537 & \textbf{0.867} & \textbf{0.735} \\
        \bottomrule
    \end{tabular}
    }
\end{table}

\begin{table}[htbp]
    \centering
    \caption{Performance comparison of different models for single-cell batch integration on the Bone Marrow dataset.}
    \label{tab:benchmark_bone_marrow}
    \resizebox{\textwidth}{!}{
    \begin{tabular}{lcccccccccccc}
        \toprule
         & \multicolumn{4}{c}{Biological conservation} & \multicolumn{5}{c}{Batch correction} & \multicolumn{3}{c}{\textbf{Total score}} \\
        \cmidrule(r){2-5} \cmidrule(r){6-10} \cmidrule{11-13}
        \textbf{Models} & NMI & ARI & ASW & cLISI & BRAS & iLISI & kBET & Conn & PCR & \textbf{Batch} & \textbf{Bio} & \textbf{Total} \\
        \midrule
        PC-HVG & 0.659 & 0.611 & 0.510 & 0.999 & 0.602 & 0.027 & 0.231 & 0.642 & 0.000 & 0.300 & 0.695 & 0.537 \\
        scVI & 0.714 & 0.556 & 0.535 & 0.999 & 0.604 & 0.042 & 0.237 & 0.766 & 0.352 & 0.400 & 0.701 & 0.581 \\
        scGPT & 0.679 & 0.487 & 0.525 & 0.998 & 0.637 & 0.025 & 0.279 & 0.685 & 0.176 & 0.360 & 0.672 & 0.547 \\
        Geneformer & 0.636 & 0.585 & 0.499 & 0.998 & 0.709 & 0.039 & 0.217 & 0.582 & 0.231 & 0.356 & 0.680 & 0.550 \\
        STATE-SE & 0.653 & 0.597 & 0.515 & 0.999 & 0.860 & 0.014 & 0.228 & 0.381 & 0.530 & 0.403 & 0.691 & 0.576 \\
        Stack & 0.733 & 0.647 & 0.507 & 0.994 & \textbf{0.866} & 0.112 & 0.296 & 0.794 & \textbf{0.658} & 0.545 & 0.720 & 0.650 \\
        \textbf{CellMSA} & \textbf{0.918} & \textbf{0.908} & \textbf{0.676} & \textbf{1.000} & 0.689 & \textbf{0.236} & \textbf{0.554} & \textbf{0.948} & 0.543 & \textbf{0.594} & \textbf{0.876} & \textbf{0.763} \\
        \bottomrule
    \end{tabular}
    }
\end{table}

\begin{table}[htbp]
    \centering
    \caption{Performance comparison of different models for single-cell batch integration on the Ear dataset.}
    \label{tab:benchmark_ear}
    \resizebox{\textwidth}{!}{
    \begin{tabular}{lcccccccccccc}
        \toprule
         & \multicolumn{4}{c}{Biological conservation} & \multicolumn{5}{c}{Batch correction} & \multicolumn{3}{c}{\textbf{Total score}} \\
        \cmidrule(r){2-5} \cmidrule(r){6-10} \cmidrule{11-13}
        \textbf{Models} & NMI & ARI & ASW & cLISI & BRAS & iLISI & kBET & Conn & PCR & \textbf{Batch} & \textbf{Bio} & \textbf{Total} \\
        \midrule
        PC-HVG & 0.760 & 0.717 & 0.546 & \textbf{1.000} & 0.638 & \textbf{0.000} & 0.802 & 0.858 & 0.000 & 0.460 & 0.756 & 0.637 \\
        scVI & 0.687 & 0.685 & 0.592 & \textbf{1.000} & 0.613 & \textbf{0.000} & \textbf{0.823} & 0.854 & 0.018 & 0.461 & 0.741 & 0.629 \\
        scGPT & 0.790 & 0.818 & 0.587 & \textbf{1.000} & 0.718 & \textbf{0.000} & 0.789 & 0.829 & 0.343 & 0.536 & 0.799 & 0.694 \\
        Geneformer & 0.790 & 0.665 & 0.538 & \textbf{1.000} & 0.829 & \textbf{0.000} & 0.809 & 0.845 & 0.596 & 0.616 & 0.748 & 0.695 \\
        STATE-SE & 0.853 & 0.823 & 0.542 & \textbf{1.000} & \textbf{0.877} & \textbf{0.000} & 0.775 & 0.592 & 0.604 & 0.570 & 0.804 & 0.711 \\
        Stack & 0.861 & 0.833 & 0.522 & 0.999 & 0.864 & \textbf{0.000} & 0.747 & 0.845 & \textbf{0.651} & \textbf{0.621} & 0.804 & \textbf{0.731} \\
        \textbf{CellMSA} & \textbf{0.952} & \textbf{0.963} & \textbf{0.717} & \textbf{1.000} & 0.633 & \textbf{0.000} & 0.765 & \textbf{0.887} & 0.000 & 0.457 & \textbf{0.908} & 0.728 \\
        \bottomrule
    \end{tabular}
    }
\end{table}

\begin{table}[htbp]
    \centering
    \caption{Performance comparison of different models for single-cell batch integration on the Eye dataset.}
    \label{tab:benchmark_eye}
    \resizebox{\textwidth}{!}{
    \begin{tabular}{lcccccccccccc}
        \toprule
         & \multicolumn{4}{c}{Biological conservation} & \multicolumn{5}{c}{Batch correction} & \multicolumn{3}{c}{\textbf{Total score}} \\
        \cmidrule(r){2-5} \cmidrule(r){6-10} \cmidrule{11-13}
        \textbf{Models} & NMI & ARI & ASW & cLISI & BRAS & iLISI & kBET & Conn & PCR & \textbf{Batch} & \textbf{Bio} & \textbf{Total} \\
        \midrule
        PC-HVG & 0.799 & 0.577 & 0.562 & \textbf{1.000} & 0.643 & 0.014 & 0.489 & 0.703 & 0.000 & 0.370 & 0.735 & 0.589 \\
        scVI & 0.823 & 0.726 & 0.555 & \textbf{1.000} & 0.662 & 0.024 & 0.426 & 0.752 & 0.382 & 0.449 & 0.776 & 0.645 \\
        scGPT & 0.795 & 0.523 & 0.565 & \textbf{1.000} & 0.652 & 0.012 & 0.421 & 0.713 & 0.000 & 0.360 & 0.721 & 0.576 \\
        Geneformer & 0.810 & 0.653 & 0.544 & \textbf{1.000} & 0.732 & 0.010 & 0.419 & 0.699 & 0.000 & 0.372 & 0.752 & 0.600 \\
        STATE-SE & 0.795 & 0.568 & 0.519 & \textbf{1.000} & \textbf{0.882} & 0.006 & 0.366 & 0.504 & 0.279 & 0.408 & 0.721 & 0.595 \\
        Stack & 0.837 & 0.756 & 0.512 & \textbf{1.000} & 0.873 & 0.044 & 0.456 & 0.771 & \textbf{0.434} & \textbf{0.516} & 0.776 & 0.672 \\
        \textbf{CellMSA} & \textbf{0.962} & \textbf{0.977} & \textbf{0.720} & \textbf{1.000} & 0.707 & \textbf{0.127} & \textbf{0.630} & \textbf{0.933} & 0.000 & 0.479 & \textbf{0.915} & \textbf{0.740} \\
        \bottomrule
    \end{tabular}
    }
\end{table}

\begin{table}[htbp]
    \centering
    \caption{Performance comparison of different models for single-cell batch integration on the Fat dataset.}
    \label{tab:benchmark_fat}
    \resizebox{\textwidth}{!}{
    \begin{tabular}{lcccccccccccc}
        \toprule
         & \multicolumn{4}{c}{Biological conservation} & \multicolumn{5}{c}{Batch correction} & \multicolumn{3}{c}{\textbf{Total score}} \\
        \cmidrule(r){2-5} \cmidrule(r){6-10} \cmidrule{11-13}
        \textbf{Models} & NMI & ARI & ASW & cLISI & BRAS & iLISI & kBET & Conn & PCR & \textbf{Batch} & \textbf{Bio} & \textbf{Total} \\
        \midrule
        PC-HVG & 0.822 & 0.800 & 0.583 & \textbf{1.000} & 0.696 & 0.022 & 0.442 & 0.830 & 0.000 & 0.398 & 0.801 & 0.640 \\
        scVI & 0.803 & 0.756 & 0.535 & \textbf{1.000} & 0.727 & 0.033 & 0.417 & 0.850 & 0.223 & 0.450 & 0.774 & 0.644 \\
        scGPT & 0.823 & 0.800 & 0.566 & \textbf{1.000} & 0.736 & 0.035 & 0.390 & 0.832 & 0.135 & 0.426 & 0.797 & 0.649 \\
        Geneformer & 0.778 & 0.700 & 0.527 & \textbf{1.000} & 0.811 & 0.026 & 0.446 & 0.779 & 0.061 & 0.425 & 0.751 & 0.621 \\
        STATE-SE & 0.778 & 0.650 & 0.520 & \textbf{1.000} & \textbf{0.911} & 0.002 & 0.321 & 0.519 & 0.318 & 0.414 & 0.737 & 0.608 \\
        Stack & 0.834 & 0.804 & 0.503 & \textbf{1.000} & 0.888 & 0.064 & 0.434 & 0.812 & \textbf{0.512} & \textbf{0.542} & 0.785 & 0.688 \\
        \textbf{CellMSA} & \textbf{0.957} & \textbf{0.916} & \textbf{0.714} & \textbf{1.000} & 0.789 & \textbf{0.212} & \textbf{0.716} & \textbf{0.937} & 0.000 & 0.531 & \textbf{0.897} & \textbf{0.750} \\
        \bottomrule
    \end{tabular}
    }
\end{table}

\begin{table}[htbp]
    \centering
    \caption{Performance comparison of different models for single-cell batch integration on the Heart dataset.}
    \label{tab:benchmark_heart}
    \resizebox{\textwidth}{!}{
    \begin{tabular}{lcccccccccccc}
        \toprule
         & \multicolumn{4}{c}{Biological conservation} & \multicolumn{5}{c}{Batch correction} & \multicolumn{3}{c}{\textbf{Total score}} \\
        \cmidrule(r){2-5} \cmidrule(r){6-10} \cmidrule{11-13}
        \textbf{Models} & NMI & ARI & ASW & cLISI & BRAS & iLISI & kBET & Conn & PCR & \textbf{Batch} & \textbf{Bio} & \textbf{Total} \\
        \midrule
        PC-HVG & 0.738 & 0.716 & 0.591 & \textbf{1.000} & 0.617 & 0.031 & 0.562 & 0.840 & 0.000 & 0.410 & 0.761 & 0.621 \\
        scVI & 0.698 & 0.517 & 0.563 & \textbf{1.000} & 0.632 & 0.019 & 0.541 & 0.840 & 0.474 & 0.501 & 0.695 & 0.617 \\
        scGPT & 0.733 & 0.648 & 0.563 & \textbf{1.000} & 0.666 & 0.046 & 0.553 & 0.862 & 0.311 & 0.488 & 0.736 & 0.637 \\
        Geneformer & 0.726 & 0.656 & 0.538 & \textbf{1.000} & 0.733 & 0.036 & 0.541 & 0.816 & 0.474 & 0.520 & 0.730 & 0.646 \\
        STATE-SE & 0.753 & 0.710 & 0.529 & \textbf{1.000} & \textbf{0.853} & 0.000 & 0.554 & 0.537 & 0.516 & 0.492 & 0.748 & 0.646 \\
        Stack & 0.748 & 0.710 & 0.512 & 0.998 & 0.837 & 0.089 & 0.598 & 0.884 & \textbf{0.651} & \textbf{0.612} & 0.742 & 0.690 \\
        \textbf{CellMSA} & \textbf{0.909} & \textbf{0.845} & \textbf{0.775} & \textbf{1.000} & 0.684 & \textbf{0.198} & \textbf{0.707} & \textbf{0.961} & 0.277 & 0.566 & \textbf{0.882} & \textbf{0.756} \\
        \bottomrule
    \end{tabular}
    }
\end{table}

\begin{table}[htbp]
    \centering
    \caption{Performance comparison of different models for single-cell batch integration on the Large Intestine dataset.}
    \label{tab:benchmark_large_intestine}
    \resizebox{\textwidth}{!}{
    \begin{tabular}{lcccccccccccc}
        \toprule
         & \multicolumn{4}{c}{Biological conservation} & \multicolumn{5}{c}{Batch correction} & \multicolumn{3}{c}{\textbf{Total score}} \\
        \cmidrule(r){2-5} \cmidrule(r){6-10} \cmidrule{11-13}
        \textbf{Models} & NMI & ARI & ASW & cLISI & BRAS & iLISI & kBET & Conn & PCR & \textbf{Batch} & \textbf{Bio} & \textbf{Total} \\
        \midrule
        PC-HVG & 0.731 & 0.470 & 0.522 & 0.999 & 0.575 & 0.015 & 0.258 & 0.784 & 0.000 & 0.326 & 0.680 & 0.539 \\
        scVI & 0.739 & 0.511 & 0.530 & 0.999 & 0.615 & 0.022 & 0.293 & 0.835 & 0.204 & 0.394 & 0.695 & 0.574 \\
        scGPT & 0.750 & 0.506 & 0.535 & 0.999 & 0.629 & 0.018 & 0.260 & 0.771 & 0.305 & 0.396 & 0.698 & 0.577 \\
        Geneformer & 0.732 & 0.515 & 0.509 & 0.998 & 0.750 & 0.032 & 0.254 & 0.747 & 0.412 & 0.439 & 0.689 & 0.589 \\
        STATE-SE & 0.769 & 0.631 & 0.519 & 0.999 & \textbf{0.890} & 0.009 & 0.247 & 0.511 & 0.548 & 0.441 & 0.730 & 0.614 \\
        Stack & 0.780 & 0.607 & 0.514 & 0.996 & 0.866 & 0.057 & 0.415 & 0.865 & \textbf{0.722} & \textbf{0.585} & 0.725 & 0.669 \\
        \textbf{CellMSA} & \textbf{0.890} & \textbf{0.821} & \textbf{0.650} & \textbf{1.000} & 0.762 & \textbf{0.212} & \textbf{0.628} & \textbf{0.978} & 0.062 & 0.528 & \textbf{0.840} & \textbf{0.715} \\
        \bottomrule
    \end{tabular}
    }
\end{table}

\begin{table}[htbp]
    \centering
    \caption{Performance comparison of different models for single-cell batch integration on the Liver dataset.}
    \label{tab:benchmark_liver}
    \resizebox{\textwidth}{!}{
    \begin{tabular}{lcccccccccccc}
        \toprule
         & \multicolumn{4}{c}{Biological conservation} & \multicolumn{5}{c}{Batch correction} & \multicolumn{3}{c}{\textbf{Total score}} \\
        \cmidrule(r){2-5} \cmidrule(r){6-10} \cmidrule{11-13}
        \textbf{Models} & NMI & ARI & ASW & cLISI & BRAS & iLISI & kBET & Conn & PCR & \textbf{Batch} & \textbf{Bio} & \textbf{Total} \\
        \midrule
        PC-HVG & 0.776 & 0.598 & 0.614 & \textbf{1.000} & 0.584 & 0.007 & 0.271 & 0.765 & 0.000 & 0.325 & 0.747 & 0.578 \\
        scVI & 0.747 & 0.469 & 0.599 & \textbf{1.000} & 0.635 & 0.019 & 0.275 & 0.786 & 0.236 & 0.390 & 0.704 & 0.578 \\
        scGPT & 0.759 & 0.591 & 0.603 & \textbf{1.000} & 0.582 & 0.003 & 0.233 & 0.767 & 0.115 & 0.340 & 0.738 & 0.579 \\
        Geneformer & 0.711 & 0.491 & 0.564 & \textbf{1.000} & 0.685 & 0.002 & 0.208 & 0.727 & 0.355 & 0.395 & 0.691 & 0.573 \\
        STATE-SE & 0.733 & 0.597 & 0.557 & \textbf{1.000} & \textbf{0.849} & 0.000 & 0.208 & 0.486 & 0.435 & 0.396 & 0.721 & 0.591 \\
        Stack & 0.790 & 0.604 & 0.522 & \textbf{1.000} & 0.841 & 0.032 & 0.236 & 0.839 & \textbf{0.679} & 0.526 & 0.729 & 0.647 \\
        \textbf{CellMSA} & \textbf{0.890} & \textbf{0.747} & \textbf{0.756} & \textbf{1.000} & 0.757 & \textbf{0.224} & \textbf{0.606} & \textbf{0.966} & 0.455 & \textbf{0.601} & \textbf{0.848} & \textbf{0.750} \\
        \bottomrule
    \end{tabular}
    }
\end{table}

\begin{table}[htbp]
    \centering
    \caption{Performance comparison of different models for single-cell batch integration on the Lung dataset.}
    \label{tab:benchmark_lung}
    \resizebox{\textwidth}{!}{
    \begin{tabular}{lcccccccccccc}
        \toprule
         & \multicolumn{4}{c}{Biological conservation} & \multicolumn{5}{c}{Batch correction} & \multicolumn{3}{c}{\textbf{Total score}} \\
        \cmidrule(r){2-5} \cmidrule(r){6-10} \cmidrule{11-13}
        \textbf{Models} & NMI & ARI & ASW & cLISI & BRAS & iLISI & kBET & Conn & PCR & \textbf{Batch} & \textbf{Bio} & \textbf{Total} \\
        \midrule
        PC-HVG & 0.758 & 0.584 & 0.595 & \textbf{1.000} & 0.643 & 0.013 & 0.167 & 0.784 & 0.000 & 0.321 & 0.734 & 0.569 \\
        scVI & 0.813 & 0.703 & 0.597 & \textbf{1.000} & 0.691 & 0.037 & 0.235 & 0.845 & 0.000 & 0.361 & 0.778 & 0.612 \\
        scGPT & 0.760 & 0.574 & 0.579 & \textbf{1.000} & 0.700 & 0.015 & 0.148 & 0.807 & 0.000 & 0.334 & 0.728 & 0.571 \\
        Geneformer & 0.735 & 0.512 & 0.539 & \textbf{1.000} & 0.748 & 0.021 & 0.169 & 0.748 & 0.268 & 0.391 & 0.697 & 0.574 \\
        STATE-SE & 0.736 & 0.451 & 0.539 & \textbf{1.000} & \textbf{0.889} & 0.002 & 0.192 & 0.419 & 0.354 & 0.371 & 0.681 & 0.557 \\
        Stack & 0.816 & 0.673 & 0.519 & \textbf{1.000} & 0.876 & 0.084 & 0.242 & 0.865 & 0.552 & 0.524 & 0.752 & 0.660 \\
        \textbf{CellMSA} & \textbf{0.895} & \textbf{0.834} & \textbf{0.675} & \textbf{1.000} & 0.787 & \textbf{0.334} & \textbf{0.520} & \textbf{0.956} & \textbf{0.588} & \textbf{0.637} & \textbf{0.851} & \textbf{0.765} \\
        \bottomrule
    \end{tabular}
    }
\end{table}

\begin{table}[htbp]
    \centering
    \caption{Performance comparison of different models for single-cell batch integration on the Lymph Node dataset.}
    \label{tab:benchmark_lymph_node}
    \resizebox{\textwidth}{!}{
    \begin{tabular}{lcccccccccccc}
        \toprule
         & \multicolumn{4}{c}{Biological conservation} & \multicolumn{5}{c}{Batch correction} & \multicolumn{3}{c}{\textbf{Total score}} \\
        \cmidrule(r){2-5} \cmidrule(r){6-10} \cmidrule{11-13}
        \textbf{Models} & NMI & ARI & ASW & cLISI & BRAS & iLISI & kBET & Conn & PCR & \textbf{Batch} & \textbf{Bio} & \textbf{Total} \\
        \midrule
        PC-HVG & 0.556 & 0.406 & 0.585 & 0.999 & 0.650 & 0.060 & 0.267 & 0.641 & 0.000 & 0.323 & 0.637 & 0.511 \\
        scVI & 0.638 & 0.507 & 0.541 & 0.999 & 0.692 & 0.089 & 0.227 & 0.690 & 0.000 & 0.339 & 0.671 & 0.539 \\
        scGPT & 0.586 & 0.426 & 0.581 & 0.999 & 0.699 & 0.070 & 0.311 & 0.682 & 0.000 & 0.352 & 0.648 & 0.530 \\
        Geneformer & 0.564 & 0.349 & 0.528 & 0.999 & 0.775 & 0.088 & 0.265 & 0.637 & 0.000 & 0.353 & 0.610 & 0.507 \\
        STATE-SE & 0.587 & 0.445 & 0.515 & 0.998 & \textbf{0.894} & 0.046 & 0.306 & 0.473 & \textbf{0.581} & 0.460 & 0.636 & 0.566 \\
        Stack & 0.726 & \textbf{0.714} & 0.501 & 0.997 & 0.887 & 0.205 & 0.207 & 0.691 & 0.514 & 0.501 & 0.734 & 0.641 \\
        \textbf{CellMSA} & \textbf{0.791} & 0.611 & \textbf{0.678} & \textbf{1.000} & 0.772 & \textbf{0.387} & \textbf{0.644} & \textbf{0.946} & 0.493 & \textbf{0.648} & \textbf{0.770} & \textbf{0.721} \\
        \bottomrule
    \end{tabular}
    }
\end{table}

\begin{table}[htbp]
    \centering
    \caption{Performance comparison of different models for single-cell batch integration on the Mammary dataset.}
    \label{tab:benchmark_mammary}
    \resizebox{\textwidth}{!}{
    \begin{tabular}{lcccccccccccc}
        \toprule
         & \multicolumn{4}{c}{Biological conservation} & \multicolumn{5}{c}{Batch correction} & \multicolumn{3}{c}{\textbf{Total score}} \\
        \cmidrule(r){2-5} \cmidrule(r){6-10} \cmidrule{11-13}
        \textbf{Models} & NMI & ARI & ASW & cLISI & BRAS & iLISI & kBET & Conn & PCR & \textbf{Batch} & \textbf{Bio} & \textbf{Total} \\
        \midrule
        PC-HVG & 0.782 & 0.529 & 0.587 & \textbf{1.000} & 0.603 & 0.000 & 0.277 & 0.847 & 0.000 & 0.346 & 0.725 & 0.573 \\
        scVI & 0.767 & 0.447 & 0.622 & \textbf{1.000} & 0.487 & 0.000 & 0.254 & 0.931 & 0.000 & 0.334 & 0.709 & 0.559 \\
        scGPT & 0.751 & 0.421 & 0.568 & \textbf{1.000} & 0.642 & 0.000 & 0.267 & 0.835 & 0.032 & 0.355 & 0.685 & 0.553 \\
        Geneformer & 0.704 & 0.335 & 0.528 & \textbf{1.000} & 0.769 & 0.000 & 0.235 & 0.759 & 0.475 & 0.448 & 0.642 & 0.564 \\
        STATE-SE & 0.740 & 0.415 & 0.528 & \textbf{1.000} & \textbf{0.866} & 0.000 & 0.216 & 0.524 & \textbf{0.515} & 0.424 & 0.671 & 0.572 \\
        Stack & 0.744 & 0.420 & 0.517 & \textbf{1.000} & 0.815 & 0.000 & 0.219 & 0.861 & 0.514 & 0.482 & 0.670 & 0.595 \\
        \textbf{CellMSA} & \textbf{0.959} & \textbf{0.972} & \textbf{0.776} & \textbf{1.000} & 0.720 & \textbf{0.419} & \textbf{0.527} & \textbf{0.949} & 0.314 & \textbf{0.586} & \textbf{0.927} & \textbf{0.790} \\
        \bottomrule
    \end{tabular}
    }
\end{table}

\begin{table}[htbp]
    \centering
    \caption{Performance comparison of different models for single-cell batch integration on the Muscle dataset.}
    \label{tab:benchmark_muscle}
    \resizebox{\textwidth}{!}{
    \begin{tabular}{lcccccccccccc}
        \toprule
         & \multicolumn{4}{c}{Biological conservation} & \multicolumn{5}{c}{Batch correction} & \multicolumn{3}{c}{\textbf{Total score}} \\
        \cmidrule(r){2-5} \cmidrule(r){6-10} \cmidrule{11-13}
        \textbf{Models} & NMI & ARI & ASW & cLISI & BRAS & iLISI & kBET & Conn & PCR & \textbf{Batch} & \textbf{Bio} & \textbf{Total} \\
        \midrule
        PC-HVG & 0.766 & 0.825 & 0.552 & \textbf{1.000} & 0.579 & 0.024 & 0.274 & 0.678 & 0.000 & 0.311 & 0.786 & 0.596 \\
        scVI & 0.806 & \textbf{0.878} & 0.560 & \textbf{1.000} & 0.657 & 0.061 & 0.336 & 0.765 & 0.025 & 0.369 & \textbf{0.811} & 0.634 \\
        scGPT & 0.729 & 0.574 & 0.561 & \textbf{1.000} & 0.622 & 0.029 & 0.283 & 0.712 & 0.000 & 0.329 & 0.716 & 0.561 \\
        Geneformer & 0.712 & 0.615 & 0.525 & \textbf{1.000} & 0.698 & 0.035 & 0.316 & 0.663 & 0.000 & 0.342 & 0.713 & 0.565 \\
        STATE-SE & 0.678 & 0.367 & 0.533 & \textbf{1.000} & \textbf{0.880} & 0.001 & 0.242 & 0.450 & 0.292 & 0.373 & 0.645 & 0.536 \\
        Stack & 0.797 & 0.871 & 0.510 & \textbf{1.000} & 0.844 & 0.100 & 0.355 & 0.755 & 0.516 & 0.514 & 0.795 & 0.682 \\
        \textbf{CellMSA} & \textbf{0.851} & 0.654 & \textbf{0.663} & \textbf{1.000} & 0.714 & \textbf{0.291} & \textbf{0.658} & \textbf{0.951} & \textbf{0.614} & \textbf{0.646} & 0.792 & \textbf{0.733} \\
        \bottomrule
    \end{tabular}
    }
\end{table}

\begin{table}[htbp]
    \centering
    \caption{Performance comparison of different models for single-cell batch integration on the Ovary dataset.}
    \label{tab:benchmark_ovary}
    \resizebox{\textwidth}{!}{
    \begin{tabular}{lcccccccccccc}
        \toprule
         & \multicolumn{4}{c}{Biological conservation} & \multicolumn{5}{c}{Batch correction} & \multicolumn{3}{c}{\textbf{Total score}} \\
        \cmidrule(r){2-5} \cmidrule(r){6-10} \cmidrule{11-13}
        \textbf{Models} & NMI & ARI & ASW & cLISI & BRAS & iLISI & kBET & Conn & PCR & \textbf{Batch} & \textbf{Bio} & \textbf{Total} \\
        \midrule
        PC-HVG & 0.565 & 0.303 & 0.594 & \textbf{1.000} & 0.653 & 0.004 & 0.374 & 0.850 & 0.000 & 0.376 & 0.616 & 0.520 \\
        scVI & 0.520 & 0.411 & 0.526 & \textbf{1.000} & 0.659 & 0.072 & 0.282 & 0.868 & 0.091 & 0.394 & 0.614 & 0.526 \\
        scGPT & 0.588 & 0.318 & 0.613 & \textbf{1.000} & 0.702 & 0.008 & 0.307 & 0.860 & 0.000 & 0.375 & 0.630 & 0.528 \\
        Geneformer & 0.542 & 0.288 & 0.555 & \textbf{1.000} & 0.776 & 0.010 & 0.290 & 0.779 & 0.000 & 0.371 & 0.596 & 0.506 \\
        STATE-SE & 0.610 & 0.322 & 0.512 & \textbf{1.000} & \textbf{0.881} & 0.000 & 0.344 & 0.662 & 0.334 & 0.444 & 0.611 & 0.544 \\
        Stack & 0.635 & \textbf{0.721} & 0.503 & \textbf{1.000} & 0.862 & 0.135 & 0.409 & 0.873 & 0.565 & 0.569 & 0.715 & 0.657 \\
        \textbf{CellMSA} & \textbf{0.765} & 0.600 & \textbf{0.667} & \textbf{1.000} & 0.741 & \textbf{0.442} & \textbf{0.710} & \textbf{0.952} & \textbf{0.759} & \textbf{0.721} & \textbf{0.758} & \textbf{0.743} \\
        \bottomrule
    \end{tabular}
    }
\end{table}

\begin{table}[htbp]
    \centering
    \caption{Performance comparison of different models for single-cell batch integration on the Pancreas dataset.}
    \label{tab:benchmark_pancreas}
    \resizebox{\textwidth}{!}{
    \begin{tabular}{lcccccccccccc}
        \toprule
         & \multicolumn{4}{c}{Biological conservation} & \multicolumn{5}{c}{Batch correction} & \multicolumn{3}{c}{\textbf{Total score}} \\
        \cmidrule(r){2-5} \cmidrule(r){6-10} \cmidrule{11-13}
        \textbf{Models} & NMI & ARI & ASW & cLISI & BRAS & iLISI & kBET & Conn & PCR & \textbf{Batch} & \textbf{Bio} & \textbf{Total} \\
        \midrule
        PC-HVG & 0.780 & 0.609 & 0.662 & \textbf{1.000} & 0.607 & 0.000 & 0.668 & 0.799 & 0.000 & 0.415 & 0.763 & 0.624 \\
        scVI & 0.764 & 0.589 & 0.595 & \textbf{1.000} & 0.651 & 0.001 & 0.666 & 0.815 & 0.494 & 0.525 & 0.737 & 0.652 \\
        scGPT & 0.768 & 0.564 & 0.610 & \textbf{1.000} & 0.704 & 0.002 & 0.670 & 0.767 & 0.477 & 0.524 & 0.736 & 0.651 \\
        Geneformer & 0.766 & 0.555 & 0.554 & \textbf{1.000} & 0.802 & 0.001 & 0.652 & 0.753 & 0.588 & 0.559 & 0.719 & 0.655 \\
        STATE-SE & 0.743 & 0.501 & 0.552 & \textbf{1.000} & 0.843 & 0.000 & 0.615 & 0.561 & 0.559 & 0.516 & 0.699 & 0.626 \\
        Stack & 0.789 & 0.620 & 0.524 & \textbf{1.000} & \textbf{0.846} & 0.014 & 0.658 & 0.802 & \textbf{0.751} & \textbf{0.614} & 0.733 & 0.686 \\
        \textbf{CellMSA} & \textbf{0.960} & \textbf{0.944} & \textbf{0.750} & \textbf{1.000} & 0.771 & \textbf{0.073} & \textbf{0.799} & \textbf{0.933} & 0.017 & 0.518 & \textbf{0.913} & \textbf{0.755} \\
        \bottomrule
    \end{tabular}
    }
\end{table}

\begin{table}[htbp]
    \centering
    \caption{Performance comparison of different models for single-cell batch integration on the Prostate dataset.}
    \label{tab:benchmark_prostate}
    \resizebox{\textwidth}{!}{
    \begin{tabular}{lcccccccccccc}
        \toprule
         & \multicolumn{4}{c}{Biological conservation} & \multicolumn{5}{c}{Batch correction} & \multicolumn{3}{c}{\textbf{Total score}} \\
        \cmidrule(r){2-5} \cmidrule(r){6-10} \cmidrule{11-13}
        \textbf{Models} & NMI & ARI & ASW & cLISI & BRAS & iLISI & kBET & Conn & PCR & \textbf{Batch} & \textbf{Bio} & \textbf{Total} \\
        \midrule
        PC-HVG & 0.641 & 0.405 & 0.573 & \textbf{1.000} & 0.595 & 0.000 & 0.345 & 0.813 & 0.000 & 0.351 & 0.655 & 0.533 \\
        scVI & 0.649 & 0.431 & 0.568 & \textbf{1.000} & 0.631 & 0.000 & 0.227 & 0.817 & 0.000 & 0.335 & 0.662 & 0.531 \\
        scGPT & 0.649 & 0.408 & 0.575 & \textbf{1.000} & 0.668 & 0.000 & 0.332 & 0.808 & 0.000 & 0.361 & 0.658 & 0.539 \\
        Geneformer & 0.626 & 0.401 & 0.527 & \textbf{1.000} & 0.726 & 0.000 & 0.284 & 0.788 & 0.000 & 0.360 & 0.639 & 0.527 \\
        STATE-SE & 0.676 & 0.428 & 0.536 & \textbf{1.000} & \textbf{0.874} & 0.000 & 0.220 & 0.508 & 0.204 & 0.361 & 0.660 & 0.541 \\
        Stack & 0.682 & 0.432 & 0.526 & \textbf{1.000} & 0.863 & 0.009 & 0.259 & 0.832 & \textbf{0.508} & 0.494 & 0.660 & 0.594 \\
        \textbf{CellMSA} & \textbf{0.890} & \textbf{0.653} & \textbf{0.787} & \textbf{1.000} & 0.769 & \textbf{0.068} & \textbf{0.607} & \textbf{0.972} & 0.465 & \textbf{0.576} & \textbf{0.832} & \textbf{0.730} \\
        \bottomrule
    \end{tabular}
    }
\end{table}

\begin{table}[htbp]
    \centering
    \caption{Performance comparison of different models for single-cell batch integration on the Salivary Gland dataset.}
    \label{tab:benchmark_salivary_gland}
    \resizebox{\textwidth}{!}{
    \begin{tabular}{lcccccccccccc}
        \toprule
         & \multicolumn{4}{c}{Biological conservation} & \multicolumn{5}{c}{Batch correction} & \multicolumn{3}{c}{\textbf{Total score}} \\
        \cmidrule(r){2-5} \cmidrule(r){6-10} \cmidrule{11-13}
        \textbf{Models} & NMI & ARI & ASW & cLISI & BRAS & iLISI & kBET & Conn & PCR & \textbf{Batch} & \textbf{Bio} & \textbf{Total} \\
        \midrule
        PC-HVG & 0.712 & 0.472 & 0.533 & \textbf{1.000} & 0.608 & 0.001 & 0.330 & 0.809 & 0.000 & 0.350 & 0.679 & 0.547 \\
        scVI & 0.716 & 0.387 & 0.542 & \textbf{1.000} & 0.618 & 0.005 & 0.355 & 0.822 & 0.245 & 0.409 & 0.661 & 0.560 \\
        scGPT & 0.701 & 0.442 & 0.543 & \textbf{1.000} & 0.656 & 0.003 & 0.339 & 0.781 & 0.094 & 0.374 & 0.671 & 0.553 \\
        Geneformer & 0.706 & 0.519 & 0.507 & \textbf{1.000} & 0.741 & 0.002 & 0.327 & 0.770 & 0.276 & 0.423 & 0.683 & 0.579 \\
        STATE-SE & 0.709 & 0.449 & 0.515 & \textbf{1.000} & \textbf{0.867} & 0.000 & 0.309 & 0.434 & 0.406 & 0.403 & 0.668 & 0.562 \\
        Stack & 0.738 & 0.513 & 0.505 & 0.999 & 0.860 & 0.022 & 0.345 & 0.823 & \textbf{0.544} & \textbf{0.519} & 0.689 & 0.621 \\
        \textbf{CellMSA} & \textbf{0.908} & \textbf{0.778} & \textbf{0.628} & \textbf{1.000} & 0.718 & \textbf{0.156} & \textbf{0.468} & \textbf{0.911} & 0.278 & 0.506 & \textbf{0.829} & \textbf{0.700} \\
        \bottomrule
    \end{tabular}
    }
\end{table}

\begin{table}[htbp]
    \centering
    \caption{Performance comparison of different models for single-cell batch integration on the Skin dataset.}
    \label{tab:benchmark_skin}
    \resizebox{\textwidth}{!}{
    \begin{tabular}{lcccccccccccc}
        \toprule
         & \multicolumn{4}{c}{Biological conservation} & \multicolumn{5}{c}{Batch correction} & \multicolumn{3}{c}{\textbf{Total score}} \\
        \cmidrule(r){2-5} \cmidrule(r){6-10} \cmidrule{11-13}
        \textbf{Models} & NMI & ARI & ASW & cLISI & BRAS & iLISI & kBET & Conn & PCR & \textbf{Batch} & \textbf{Bio} & \textbf{Total} \\
        \midrule
        PC-HVG & 0.778 & 0.719 & 0.542 & \textbf{1.000} & 0.677 & 0.194 & 0.261 & 0.759 & 0.000 & 0.378 & 0.760 & 0.607 \\
        scVI & 0.838 & 0.800 & 0.525 & \textbf{1.000} & 0.707 & 0.225 & 0.397 & 0.771 & 0.356 & 0.491 & 0.791 & 0.671 \\
        scGPT & 0.783 & 0.720 & 0.551 & \textbf{1.000} & 0.697 & 0.171 & 0.242 & 0.759 & 0.024 & 0.379 & 0.763 & 0.610 \\
        Geneformer & 0.757 & 0.719 & 0.509 & \textbf{1.000} & 0.753 & 0.187 & 0.243 & 0.661 & 0.000 & 0.369 & 0.746 & 0.595 \\
        STATE-SE & 0.835 & 0.816 & 0.525 & \textbf{1.000} & \textbf{0.912} & 0.102 & 0.227 & 0.624 & 0.559 & 0.485 & 0.794 & 0.670 \\
        Stack & 0.842 & 0.807 & 0.515 & 0.999 & 0.907 & 0.313 & 0.415 & 0.823 & 0.717 & 0.635 & 0.791 & 0.729 \\
        \textbf{CellMSA} & \textbf{0.911} & \textbf{0.841} & \textbf{0.671} & \textbf{1.000} & 0.756 & \textbf{0.411} & \textbf{0.652} & \textbf{0.916} & \textbf{0.757} & \textbf{0.699} & \textbf{0.856} & \textbf{0.793} \\
        \bottomrule
    \end{tabular}
    }
\end{table}

\begin{table}[htbp]
    \centering
    \caption{Performance comparison of different models for single-cell batch integration on the Small Intestine dataset.}
    \label{tab:benchmark_small_intestine}
    \resizebox{\textwidth}{!}{
    \begin{tabular}{lcccccccccccc}
        \toprule
         & \multicolumn{4}{c}{Biological conservation} & \multicolumn{5}{c}{Batch correction} & \multicolumn{3}{c}{\textbf{Total score}} \\
        \cmidrule(r){2-5} \cmidrule(r){6-10} \cmidrule{11-13}
        \textbf{Models} & NMI & ARI & ASW & cLISI & BRAS & iLISI & kBET & Conn & PCR & \textbf{Batch} & \textbf{Bio} & \textbf{Total} \\
        \midrule
        PC-HVG & 0.717 & \textbf{0.669} & 0.521 & 0.998 & 0.662 & 0.027 & 0.384 & 0.746 & 0.000 & 0.364 & 0.726 & 0.581 \\
        scVI & 0.710 & \textbf{0.669} & 0.545 & 0.999 & 0.703 & 0.061 & 0.378 & 0.821 & 0.345 & 0.461 & \textbf{0.731} & 0.623 \\
        scGPT & 0.659 & 0.532 & 0.526 & 0.997 & 0.723 & 0.084 & 0.334 & 0.773 & 0.317 & 0.446 & 0.679 & 0.586 \\
        Geneformer & 0.676 & 0.552 & 0.510 & 0.998 & 0.756 & 0.049 & 0.328 & 0.715 & 0.526 & 0.475 & 0.684 & 0.600 \\
        STATE-SE & 0.693 & 0.498 & 0.514 & \textbf{1.000} & \textbf{0.885} & 0.001 & 0.304 & 0.477 & 0.442 & 0.422 & 0.676 & 0.574 \\
        Stack & 0.623 & 0.401 & 0.510 & 0.992 & 0.872 & 0.150 & 0.400 & 0.853 & \textbf{0.773} & \textbf{0.610} & 0.632 & 0.623 \\
        \textbf{CellMSA} & \textbf{0.768} & 0.554 & \textbf{0.570} & 0.997 & 0.792 & \textbf{0.274} & \textbf{0.612} & \textbf{0.931} & 0.120 & 0.546 & 0.722 & \textbf{0.652} \\
        \bottomrule
    \end{tabular}
    }
\end{table}

\begin{table}[htbp]
    \centering
    \caption{Performance comparison of different models for single-cell batch integration on the Spleen dataset.}
    \label{tab:benchmark_spleen}
    \resizebox{\textwidth}{!}{
    \begin{tabular}{lcccccccccccc}
        \toprule
         & \multicolumn{4}{c}{Biological conservation} & \multicolumn{5}{c}{Batch correction} & \multicolumn{3}{c}{\textbf{Total score}} \\
        \cmidrule(r){2-5} \cmidrule(r){6-10} \cmidrule{11-13}
        \textbf{Models} & NMI & ARI & ASW & cLISI & BRAS & iLISI & kBET & Conn & PCR & \textbf{Batch} & \textbf{Bio} & \textbf{Total} \\
        \midrule
        PC-HVG & 0.716 & 0.617 & 0.591 & \textbf{1.000} & 0.656 & 0.111 & 0.189 & 0.734 & 0.000 & 0.338 & 0.731 & 0.574 \\
        scVI & 0.793 & 0.716 & 0.572 & \textbf{1.000} & 0.709 & 0.122 & 0.188 & 0.826 & 0.000 & 0.369 & 0.770 & 0.610 \\
        scGPT & 0.742 & 0.622 & 0.592 & \textbf{1.000} & 0.730 & 0.136 & 0.150 & 0.788 & 0.000 & 0.361 & 0.739 & 0.588 \\
        Geneformer & 0.704 & 0.551 & 0.531 & \textbf{1.000} & 0.777 & 0.155 & 0.154 & 0.728 & 0.000 & 0.363 & 0.697 & 0.563 \\
        STATE-SE & 0.727 & 0.604 & 0.531 & \textbf{1.000} & \textbf{0.898} & 0.048 & 0.175 & 0.514 & 0.439 & 0.415 & 0.716 & 0.595 \\
        Stack & 0.781 & 0.742 & 0.514 & \textbf{1.000} & 0.893 & 0.291 & 0.217 & 0.840 & 0.461 & 0.541 & 0.759 & 0.672 \\
        \textbf{CellMSA} & \textbf{0.916} & \textbf{0.887} & \textbf{0.701} & \textbf{1.000} & 0.809 & \textbf{0.434} & \textbf{0.547} & \textbf{0.927} & \textbf{0.626} & \textbf{0.669} & \textbf{0.876} & \textbf{0.793} \\
        \bottomrule
    \end{tabular}
    }
\end{table}

\begin{table}[htbp]
    \centering
    \caption{Performance comparison of different models for single-cell batch integration on the Stomach dataset.}
    \label{tab:benchmark_stomach}
    \resizebox{\textwidth}{!}{
    \begin{tabular}{lcccccccccccc}
        \toprule
         & \multicolumn{4}{c}{Biological conservation} & \multicolumn{5}{c}{Batch correction} & \multicolumn{3}{c}{\textbf{Total score}} \\
        \cmidrule(r){2-5} \cmidrule(r){6-10} \cmidrule{11-13}
        \textbf{Models} & NMI & ARI & ASW & cLISI & BRAS & iLISI & kBET & Conn & PCR & \textbf{Batch} & \textbf{Bio} & \textbf{Total} \\
        \midrule
        PC-HVG & 0.879 & 0.878 & 0.608 & \textbf{1.000} & 0.610 & 0.006 & 0.521 & 0.850 & 0.000 & 0.397 & 0.841 & 0.664 \\
        scVI & 0.856 & 0.791 & 0.586 & \textbf{1.000} & 0.646 & 0.019 & 0.499 & 0.896 & 0.277 & 0.467 & 0.808 & 0.672 \\
        scGPT & 0.847 & 0.800 & 0.592 & \textbf{1.000} & 0.628 & 0.007 & 0.486 & 0.841 & 0.189 & 0.430 & 0.810 & 0.658 \\
        Geneformer & 0.846 & 0.792 & 0.549 & \textbf{1.000} & 0.756 & 0.006 & 0.512 & 0.797 & 0.314 & 0.477 & 0.797 & 0.669 \\
        STATE-SE & 0.852 & 0.816 & 0.532 & \textbf{1.000} & \textbf{0.896} & 0.000 & 0.432 & 0.538 & 0.441 & 0.461 & 0.800 & 0.665 \\
        Stack & 0.884 & 0.891 & 0.518 & \textbf{1.000} & 0.872 & 0.032 & 0.440 & 0.889 & \textbf{0.572} & \textbf{0.561} & 0.823 & 0.718 \\
        \textbf{CellMSA} & \textbf{0.984} & \textbf{0.983} & \textbf{0.770} & \textbf{1.000} & 0.662 & \textbf{0.087} & \textbf{0.633} & \textbf{0.965} & 0.000 & 0.469 & \textbf{0.934} & \textbf{0.748} \\
        \bottomrule
    \end{tabular}
    }
\end{table}

\begin{table}[htbp]
    \centering
    \caption{Performance comparison of different models for single-cell batch integration on the Thymus dataset.}
    \label{tab:benchmark_thymus}
    \resizebox{\textwidth}{!}{
    \begin{tabular}{lcccccccccccc}
        \toprule
         & \multicolumn{4}{c}{Biological conservation} & \multicolumn{5}{c}{Batch correction} & \multicolumn{3}{c}{\textbf{Total score}} \\
        \cmidrule(r){2-5} \cmidrule(r){6-10} \cmidrule{11-13}
        \textbf{Models} & NMI & ARI & ASW & cLISI & BRAS & iLISI & kBET & Conn & PCR & \textbf{Batch} & \textbf{Bio} & \textbf{Total} \\
        \midrule
        PC-HVG & 0.673 & 0.404 & 0.529 & 0.995 & 0.544 & 0.011 & 0.262 & 0.800 & 0.000 & 0.324 & 0.650 & 0.520 \\
        scVI & 0.722 & 0.459 & 0.533 & 0.997 & 0.621 & 0.035 & 0.310 & 0.810 & 0.000 & 0.355 & 0.678 & 0.549 \\
        scGPT & 0.697 & 0.438 & 0.540 & 0.996 & 0.658 & 0.029 & 0.278 & 0.828 & 0.000 & 0.359 & 0.668 & 0.544 \\
        Geneformer & 0.690 & 0.436 & 0.517 & 0.996 & 0.731 & 0.022 & 0.280 & 0.774 & 0.000 & 0.361 & 0.660 & 0.540 \\
        STATE-SE & 0.730 & 0.496 & 0.518 & 0.998 & \textbf{0.864} & 0.005 & 0.257 & 0.492 & 0.147 & 0.353 & 0.686 & 0.553 \\
        Stack & 0.759 & 0.476 & 0.511 & 0.994 & 0.845 & 0.051 & 0.337 & 0.869 & \textbf{0.307} & 0.482 & 0.685 & 0.604 \\
        \textbf{CellMSA} & \textbf{0.837} & \textbf{0.648} & \textbf{0.626} & \textbf{0.999} & 0.755 & \textbf{0.256} & \textbf{0.663} & \textbf{0.945} & 0.000 & \textbf{0.524} & \textbf{0.777} & \textbf{0.676} \\
        \bottomrule
    \end{tabular}
    }
\end{table}

\begin{table}[htbp]
    \centering
    \caption{Performance comparison of different models for single-cell batch integration on the Tongue dataset.}
    \label{tab:benchmark_tongue}
    \resizebox{\textwidth}{!}{
    \begin{tabular}{lcccccccccccc}
        \toprule
         & \multicolumn{4}{c}{Biological conservation} & \multicolumn{5}{c}{Batch correction} & \multicolumn{3}{c}{\textbf{Total score}} \\
        \cmidrule(r){2-5} \cmidrule(r){6-10} \cmidrule{11-13}
        \textbf{Models} & NMI & ARI & ASW & cLISI & BRAS & iLISI & kBET & Conn & PCR & \textbf{Batch} & \textbf{Bio} & \textbf{Total} \\
        \midrule
        PC-HVG & 0.592 & 0.318 & 0.576 & \textbf{1.000} & 0.543 & 0.013 & 0.428 & 0.789 & 0.000 & 0.355 & 0.621 & 0.515 \\
        scVI & 0.562 & 0.204 & 0.550 & \textbf{1.000} & 0.568 & 0.034 & 0.362 & 0.851 & 0.073 & 0.378 & 0.579 & 0.498 \\
        scGPT & 0.635 & 0.352 & 0.533 & \textbf{1.000} & 0.582 & 0.015 & 0.361 & 0.807 & 0.000 & 0.353 & 0.630 & 0.519 \\
        Geneformer & 0.532 & 0.168 & 0.523 & 0.999 & 0.684 & 0.021 & 0.386 & 0.793 & 0.000 & 0.377 & 0.556 & 0.484 \\
        STATE-SE & 0.548 & 0.199 & 0.534 & \textbf{1.000} & \textbf{0.849} & 0.002 & 0.333 & 0.516 & 0.312 & 0.402 & 0.570 & 0.503 \\
        Stack & 0.599 & 0.380 & 0.513 & 0.998 & 0.832 & 0.071 & 0.422 & 0.848 & \textbf{0.422} & 0.519 & 0.622 & 0.581 \\
        \textbf{CellMSA} & \textbf{0.773} & \textbf{0.580} & \textbf{0.656} & \textbf{1.000} & 0.696 & \textbf{0.260} & \textbf{0.551} & \textbf{0.957} & 0.156 & \textbf{0.524} & \textbf{0.752} & \textbf{0.661} \\
        \bottomrule
    \end{tabular}
    }
\end{table}

\begin{table}[htbp]
    \centering
    \caption{Performance comparison of different models for single-cell batch integration on the Trachea dataset.}
    \label{tab:benchmark_trachea}
    \resizebox{\textwidth}{!}{
    \begin{tabular}{lcccccccccccc}
        \toprule
         & \multicolumn{4}{c}{Biological conservation} & \multicolumn{5}{c}{Batch correction} & \multicolumn{3}{c}{\textbf{Total score}} \\
        \cmidrule(r){2-5} \cmidrule(r){6-10} \cmidrule{11-13}
        \textbf{Models} & NMI & ARI & ASW & cLISI & BRAS & iLISI & kBET & Conn & PCR & \textbf{Batch} & \textbf{Bio} & \textbf{Total} \\
        \midrule
        PC-HVG & 0.788 & 0.653 & 0.597 & \textbf{1.000} & 0.618 & 0.002 & 0.444 & 0.780 & 0.000 & 0.369 & 0.759 & 0.603 \\
        scVI & 0.813 & 0.665 & 0.597 & \textbf{1.000} & 0.629 & 0.002 & 0.479 & 0.836 & 0.103 & 0.410 & 0.769 & 0.625 \\
        scGPT & 0.796 & 0.652 & 0.587 & \textbf{1.000} & 0.678 & 0.002 & 0.414 & 0.834 & 0.109 & 0.407 & 0.759 & 0.618 \\
        Geneformer & 0.777 & 0.625 & 0.531 & \textbf{1.000} & 0.769 & 0.001 & 0.446 & 0.771 & 0.096 & 0.417 & 0.733 & 0.607 \\
        STATE-SE & 0.830 & 0.763 & 0.531 & \textbf{1.000} & \textbf{0.876} & 0.000 & 0.361 & 0.571 & 0.346 & 0.431 & 0.781 & 0.641 \\
        Stack & 0.870 & \textbf{0.916} & 0.505 & \textbf{1.000} & 0.869 & 0.017 & 0.448 & 0.890 & \textbf{0.611} & \textbf{0.567} & 0.823 & \textbf{0.720} \\
        \textbf{CellMSA} & \textbf{0.908} & 0.711 & \textbf{0.751} & \textbf{1.000} & 0.766 & \textbf{0.042} & \textbf{0.787} & \textbf{0.986} & 0.000 & 0.516 & \textbf{0.843} & 0.712 \\
        \bottomrule
    \end{tabular}
    }
\end{table}

\begin{table}[htbp]
    \centering
    \caption{Performance comparison of different models for single-cell batch integration on the Uterus dataset.}
    \label{tab:benchmark_uterus}
    \resizebox{\textwidth}{!}{
    \begin{tabular}{lcccccccccccc}
        \toprule
         & \multicolumn{4}{c}{Biological conservation} & \multicolumn{5}{c}{Batch correction} & \multicolumn{3}{c}{\textbf{Total score}} \\
        \cmidrule(r){2-5} \cmidrule(r){6-10} \cmidrule{11-13}
        \textbf{Models} & NMI & ARI & ASW & cLISI & BRAS & iLISI & kBET & Conn & PCR & \textbf{Batch} & \textbf{Bio} & \textbf{Total} \\
        \midrule
        PC-HVG & 0.772 & 0.646 & 0.500 & \textbf{1.000} & 0.665 & 0.008 & 0.329 & 0.780 & 0.000 & 0.356 & 0.729 & 0.580 \\
        scVI & 0.742 & 0.531 & 0.500 & \textbf{1.000} & 0.694 & 0.006 & 0.276 & 0.760 & 0.000 & 0.347 & 0.693 & 0.555 \\
        scGPT & 0.756 & 0.621 & 0.510 & \textbf{1.000} & 0.693 & 0.010 & 0.304 & 0.730 & 0.000 & 0.347 & 0.722 & 0.572 \\
        Geneformer & 0.754 & 0.583 & 0.490 & \textbf{1.000} & 0.774 & 0.055 & 0.320 & 0.700 & 0.129 & 0.396 & 0.707 & 0.582 \\
        STATE-SE & 0.732 & 0.533 & 0.506 & \textbf{1.000} & \textbf{0.918} & 0.004 & 0.219 & 0.558 & 0.263 & 0.392 & 0.693 & 0.573 \\
        Stack & 0.804 & 0.769 & 0.499 & 0.999 & 0.896 & 0.098 & 0.350 & 0.782 & \textbf{0.302} & 0.486 & 0.768 & 0.655 \\
        \textbf{CellMSA} & \textbf{0.943} & \textbf{0.936} & \textbf{0.628} & \textbf{1.000} & 0.711 & \textbf{0.351} & \textbf{0.477} & \textbf{0.923} & 0.250 & \textbf{0.542} & \textbf{0.877} & \textbf{0.743} \\
        \bottomrule
    \end{tabular}
    }
\end{table}

\begin{table}[htbp]
    \centering
    \caption{Performance comparison of different models for single-cell batch integration on the Vasculature dataset.}
    \label{tab:benchmark_vasculature}
    \resizebox{\textwidth}{!}{
    \begin{tabular}{lcccccccccccc}
        \toprule
         & \multicolumn{4}{c}{Biological conservation} & \multicolumn{5}{c}{Batch correction} & \multicolumn{3}{c}{\textbf{Total score}} \\
        \cmidrule(r){2-5} \cmidrule(r){6-10} \cmidrule{11-13}
        \textbf{Models} & NMI & ARI & ASW & cLISI & BRAS & iLISI & kBET & Conn & PCR & \textbf{Batch} & \textbf{Bio} & \textbf{Total} \\
        \midrule
        PC-HVG & 0.749 & 0.688 & 0.585 & \textbf{1.000} & 0.602 & 0.075 & 0.387 & 0.810 & 0.000 & 0.375 & 0.755 & 0.603 \\
        scVI & 0.783 & 0.722 & 0.551 & \textbf{1.000} & 0.655 & 0.129 & 0.385 & 0.855 & 0.319 & 0.469 & 0.764 & 0.646 \\
        scGPT & 0.731 & 0.649 & 0.561 & \textbf{1.000} & 0.650 & 0.072 & 0.371 & 0.801 & 0.242 & 0.427 & 0.735 & 0.612 \\
        Geneformer & 0.684 & 0.503 & 0.517 & \textbf{1.000} & 0.721 & 0.060 & 0.379 & 0.728 & 0.240 & 0.426 & 0.676 & 0.576 \\
        STATE-SE & 0.791 & 0.738 & 0.524 & \textbf{1.000} & \textbf{0.881} & 0.007 & 0.299 & 0.436 & 0.458 & 0.416 & 0.763 & 0.624 \\
        Stack & 0.782 & 0.717 & 0.514 & 0.999 & \textbf{0.881} & 0.137 & 0.389 & 0.849 & \textbf{0.634} & \textbf{0.578} & 0.753 & 0.683 \\
        \textbf{CellMSA} & \textbf{0.933} & \textbf{0.924} & \textbf{0.697} & \textbf{1.000} & 0.774 & \textbf{0.276} & \textbf{0.634} & \textbf{0.960} & 0.007 & 0.530 & \textbf{0.889} & \textbf{0.745} \\
        \bottomrule
    \end{tabular}
    }
\end{table}

\FloatBarrier

\subsubsection{Label-free batch integration on Bladder}
\label{app:label_free_integration}
To evaluate integration when cell-type annotations are unavailable, we additionally use expression-based HVG--PCA--KNN context retrieval on Tabula Sapiens-Bladder, following the nearest-neighbor strategy used in the classification experiments (Appendix~\ref{app:downstream_setting}). No cell-type labels are accessed for context construction in this setting; biological labels are used only to calculate evaluation metrics. We compare CellMSA and Stack under the same label-free retrieval protocol and report scGPT and Geneformer as non-contextual baselines. For reference, we also report the label-informed CellMSA and Stack results, which use cell-type annotations for context retrieval.

\begin{table}[htbp]
\centering
\caption{Label-free and label-informed integration on Tabula Sapiens-Bladder. Higher scores are better. Comparisons within each group use the stated cell-type-label access; scores across groups should be interpreted with this distinction.}
\label{tab:label_free_integration}
\begin{tabular}{lccc}
\toprule
Method & Batch $\uparrow$ & Bio $\uparrow$ & Total $\uparrow$ \\
\midrule
\multicolumn{4}{l}{\textit{Label-free representation extraction and context retrieval}} \\
scGPT & 0.413 & 0.720 & 0.598 \\
Geneformer & 0.438 & 0.728 & 0.612 \\
Stack & 0.451 & 0.699 & 0.600 \\
CellMSA & \textbf{0.509} & \textbf{0.807} & \textbf{0.688} \\
\midrule
\multicolumn{4}{l}{\textit{Label-informed context retrieval}} \\
Stack & 0.533 & 0.808 & 0.698 \\
CellMSA & \textbf{0.571} & \textbf{0.867} & \textbf{0.748} \\
\bottomrule
\end{tabular}
\end{table}

As shown in Table~\ref{tab:label_free_integration}, CellMSA achieves a total score of 0.688 without label-informed retrieval, compared with 0.600 for label-free Stack. Cell-type annotations further improve CellMSA's total score to 0.748. These results show both that CellMSA retains an advantage among the evaluated label-free methods and that annotation-informed retrieval provides an additional benefit. This experiment is restricted to Bladder and does not isolate the contribution of individual architectural components.
\FloatBarrier

\subsection{More interpretability experiments}
\label{app:interpretability}
\subsubsection{Formulation of cell-type and condition-level gene pair representations}

To examine whether the learned CellMSA pairwise representation captures biologically meaningful programs, we first compared the gene pair representations between two representative kidney cell types, proximal tubule cells (PT) and endothelial cells (EC). For each query cell, CellMSA returns an 8-channel pairwise tensor
\begin{equation}
    \mathbf{P}_{i}^{(h)} \in \mathbb{R}^{G \times G}, \quad h = 1,\dots,8,
\end{equation}
where $G$ is the fixed gene set and $h$ denotes the pairwise channel/head. We treated each head as a distinct directed pairwise activation mode. 

For a cell type $c$, condition $s \in \{\mathrm{healthy}, \mathrm{disease}\}$, and head $h$, we computed the condition-level mean matrix
\begin{equation}
    \bar{\mathbf{P}}_{c,s}^{(h)}
    =
    \frac{1}{|\mathcal{I}_{c,s}|}
    \sum_{i \in \mathcal{I}_{c,s}} \mathbf{P}_{i}^{(h)},
\end{equation}
where $\mathcal{I}_{c,s}$ denotes the sampled cells of cell type $c$ and condition $s$. Disease-associated remodeling was quantified as
\begin{equation}
    \Delta \mathbf{P}_{c}^{(h)}
    =
    \bar{\mathbf{P}}_{c,\mathrm{disease}}^{(h)}
    -
    \bar{\mathbf{P}}_{c,\mathrm{healthy}}^{(h)}.
\end{equation}

\subsubsection{Evaluating cell-type separation and head-specific variations}

\begin{figure}[ht]
\centering
\includegraphics[width=\textwidth]{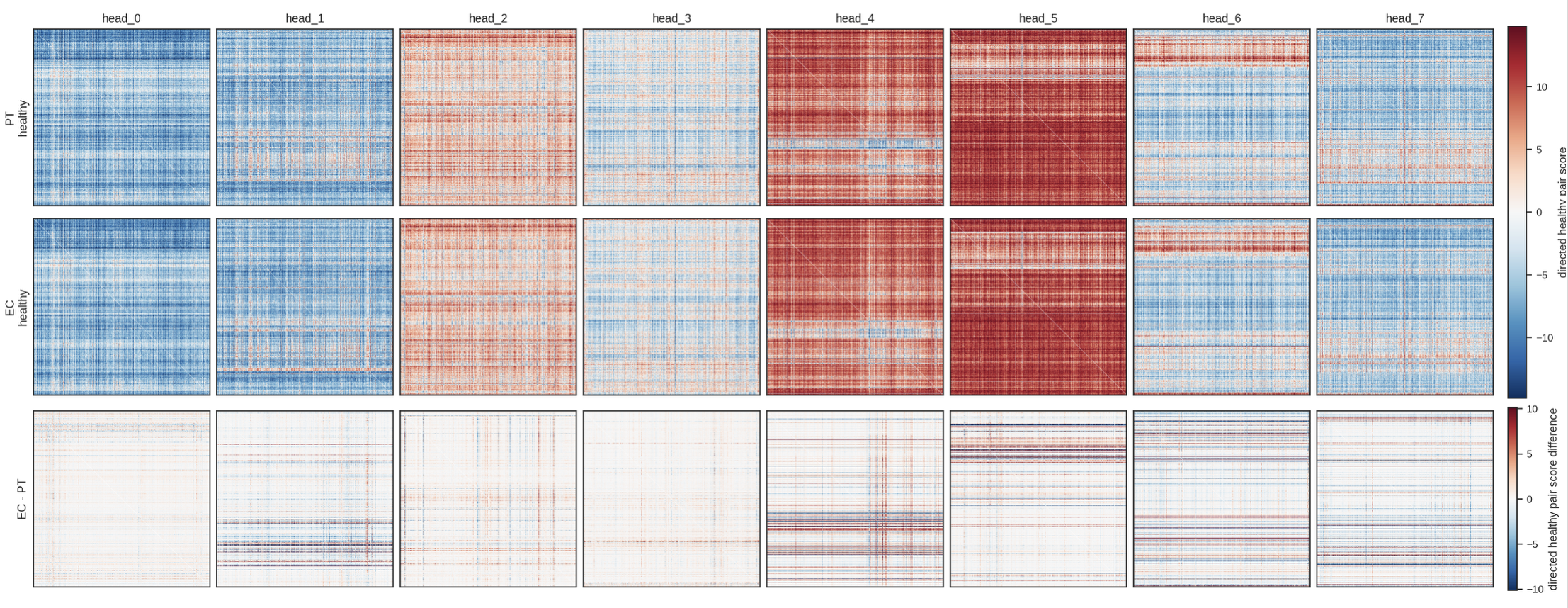}
\caption{Cell-type comparison of CellMSA gene pair representation patterns. Columns correspond to the eight CellMSA heads. Rows show PT healthy, EC healthy, and EC-minus-PT healthy pairwise matrices. Within each head, the three panels share the same gene order.}
\label{fig:figb2-1}
\end{figure}

For the cell-type comparison, we focused on healthy cells only. Figure~\ref{fig:figb2-1} summarizes this comparison. The heatmaps show that PT and EC share a broad global activation architecture but differ in head-specific block structures. This is especially visible in the EC-minus-PT row, where several heads show coherent positive or negative modules rather than diffuse random noise. 

\begin{table}[ht]
\centering
\caption{Healthy PT-vs-EC similarity of CellMSA pairwise matrices.}
\label{tab:pt-ec-healthy-pearson}
\begin{tabular}{lrrrr}
\toprule
Head & Pearson $r$ & Distance $1-r$ & Mean absolute difference & RMSE difference \\
\midrule
0 & 0.964 & 0.036 & 0.603 & 1.045 \\
1 & 0.916 & 0.084 & 1.037 & 2.073 \\
2 & 0.966 & 0.034 & 0.661 & 1.224 \\
3 & 0.974 & 0.026 & 0.552 & 0.989 \\
4 & 0.828 & 0.172 & 1.588 & 3.170 \\
5 & 0.827 & 0.173 & 1.140 & 2.403 \\
6 & 0.845 & 0.155 & 1.563 & 2.983 \\
7 & 0.869 & 0.131 & 1.188 & 2.449 \\
\midrule
Mean & 0.899 & 0.101 & 1.042 & 2.042 \\
\bottomrule
\end{tabular}
\end{table}

\begin{figure}[ht]
\centering
\includegraphics[width=0.8\textwidth]{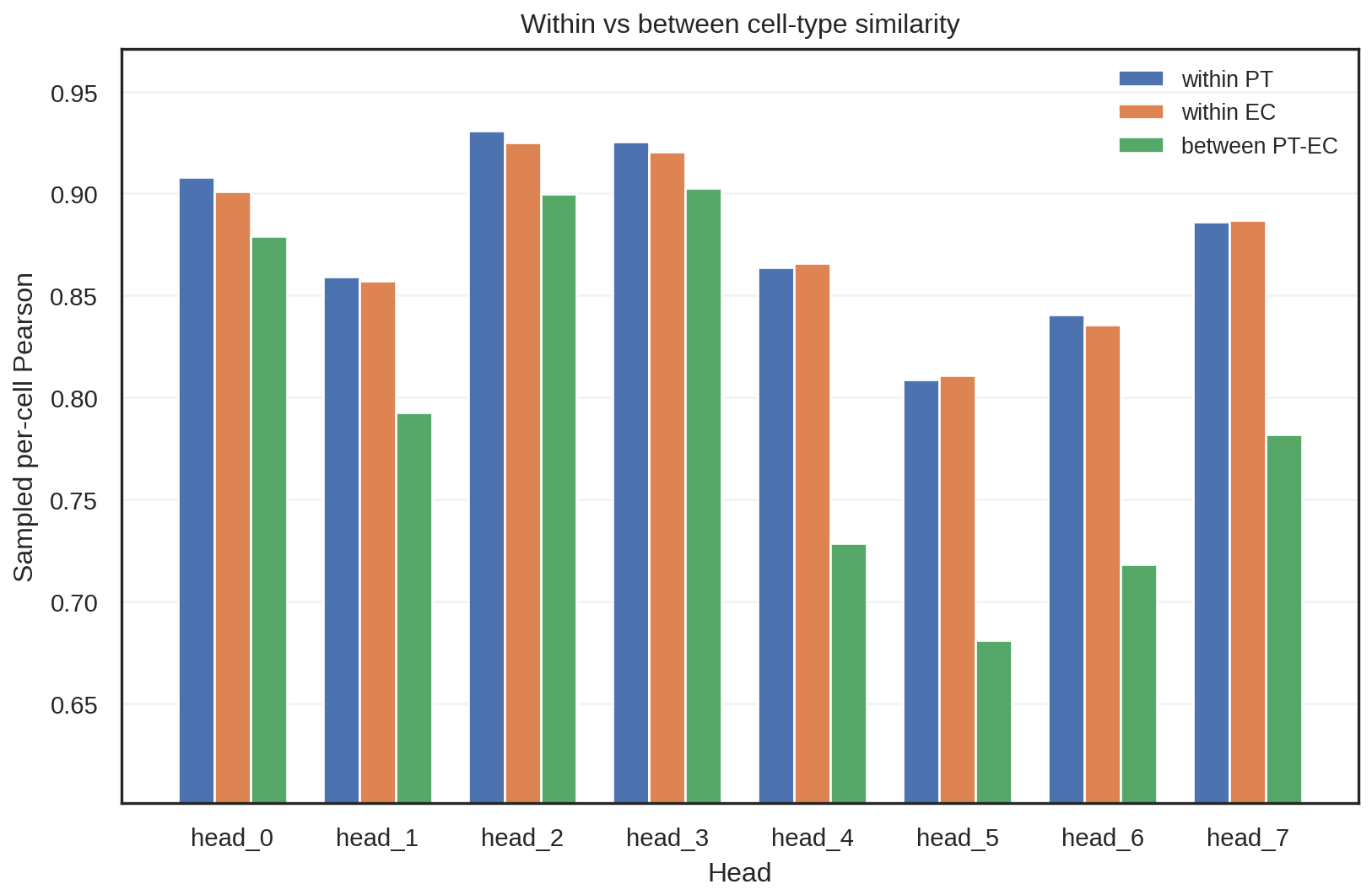}
\caption{Per-cell directed pairwise Pearson similarity shows higher within-cell-type similarity than PT-EC between-cell-type similarity across heads.}
\label{fig:figb2-2}
\end{figure}

We quantified this visual separation using Pearson correlation. The PT-vs-EC healthy Pearson similarity ranged from $r=0.827$ to $r=0.974$ across heads, with mean $r=0.899$. This indicates that the learned directed matrices retain a shared global structure, while the magnitude of cell-type separation is head-dependent. Head 3 was the most similar between PT and EC ($r=0.974$), whereas head 5 showed the lowest Pearson similarity ($r=0.827$). We then compute within-cell-type and between-cell-type similarity to investigate whether matrices from the same cell type are more similar to one another than to matrices from other cell types. As shown in Figure \ref{fig:figb2-2}, the within-cell-type Pearson similarity was consistently higher than the PT-EC between-cell-type similarity across all heads. Averaged over heads and cell types, the within-cell-type Pearson was 0.877, compared with a between-cell-type Pearson of 0.798, giving a mean separation margin of $\Delta r=0.079$. The separation was head-dependent: heads 4, 5, 6, and 7 showed the largest margins, with average $\Delta r$ values of 0.136, 0.129, 0.120, and 0.105, respectively, whereas heads 0, 2, and 3 showed smaller but still positive margins (Table~\ref{tab:pt-ec-healthy-pearson}). These results indicate that CellMSA pairwise patterns show measurable cell-type-specific deviations, particularly in the heads that showed larger PT-vs-EC differences.

To identify genes contributing to the PT-vs-EC difference in healthy pairwise activation, we summarized each gene's cell-type difference across all heads. For gene $g$ and head $h$, we computed the mean absolute PT-vs-EC pairwise difference
\begin{equation}
    D_{h}^{\mathrm{PT,EC}}(g)
    =
    \frac{1}{G-1}
    \sum_{j \neq g}
    \left|
    \bar{P}_{\mathrm{EC},\mathrm{healthy}}^{(h)}(g,j)
    -
    \bar{P}_{\mathrm{PT},\mathrm{healthy}}^{(h)}(g,j)
    \right|,
\end{equation}
and then averaged this score across heads. The top-ranked directed PT-vs-EC difference genes included MEIS2, ST6GALNAC3, LRP2, TEK, BICC1, ACSM2A, EMCN, PECAM1, GNA14, FLT1, PLAT, PTPRB, ZEB1, TRABD2B, ADGRF5, KALRN, CUBN, TMTC1, EPAS1, and SH3RF3. This list contains endothelial-associated genes and proximal tubule-associated genes. Therefore, the genes driving the pairwise matrix differences recover known cell identity markers, supporting the biological interpretability of the cell-type-specific activation patterns.

\begin{figure}[ht]
\centering
\includegraphics[width=1\textwidth]{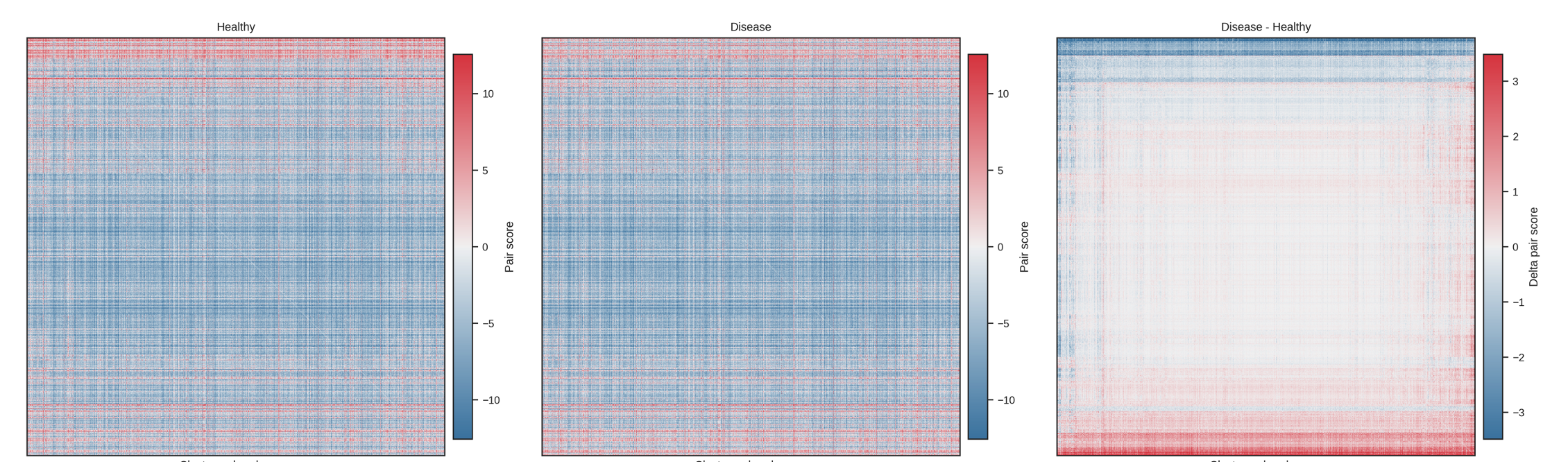}
\caption{Gene pair patterns in healthy and disease conditions within head 7.}
\label{fig:figb2-3}
\end{figure}

\subsubsection{Evaluating disease-driven alterations in gene pair representations}

We next investigated how these gene-gene relationships are reconfigured in the disease state. To identify key regulatory genes driving disease-associated network within a given cell type, we summarized each gene's hub strength in the CellMSA pairwise matrix. For gene $g$ in cell type $c$, condition $s$, and head $h$, we defined the hub score and hub strength as
\begin{equation}
    H_{c,s,h}(g)
    =
    \frac{1}{G-1}
    \sum_{j \neq g}
    \bar{P}_{c,s}^{(h)}(g,j),
\end{equation}
We then computed disease-associated changes in both quantities:
\begin{equation}
    \Delta H_{c,h}(g)
    =
    H_{c,\mathrm{disease},h}^{\mathrm{signed}}(g)
    -
    H_{c,\mathrm{healthy},h}^{\mathrm{signed}}(g),
\end{equation}

In addition, to capture genes whose pairwise partners were substantially remodeled even if the total strength did not monotonically increase or decrease, we computed a delta-hub score
\begin{equation}
    H_{c,h}^{\Delta \mathrm{abs}}(g)
    =
    \frac{1}{G-1}
    \sum_{j \neq g}
    \left|
    \bar{P}_{c,\mathrm{disease}}^{(h)}(g,j)
    -
    \bar{P}_{c,\mathrm{healthy}}^{(h)}(g,j)
    \right|.
\end{equation}
Genes with large $H_{c,h}^{\Delta \mathrm{abs}}$ are interpreted as genes whose pairwise activation neighborhoods are most remodeled by disease. 
We next examined whether different CellMSA heads encode disease-associated pairwise activation modes. We used the head-level delta score to prioritize heads for downstream interpretation:
\begin{equation}
    \frac{1}{|\mathcal{E}|}
    \sum_{(g,j) \in \mathcal{E}}
    \left|
    \Delta P_{c}^{(h)}(g,j)
    \right|,
\end{equation}
For the PT analysis, the strongest remodeled heads were head 4, head 6, and head 7, with mean absolute pairwise delta scores of 0.529, 0.512, and 0.380, respectively. Head 7 showed a balanced signed delta score ($\mathrm{mean}(\Delta P)=-7.68\times 10^{-4}$) but substantial remodeling magnitude ($\mathrm{mean}(|\Delta P|)=0.380$; maximum $|\Delta P|=8.907$). The healthy and disease panels show the baseline pairwise activation architecture within each condition, while the delta panel highlights disease-associated rewiring, which displayed disease-remodeled gene modules in head 7 (Fig.~\ref{fig:figb2-3}). This supports that different heads capture different biological programs, with head 7 specifically capturing disease-associated patterns.

\subsubsection{External validation against STRING functional associations}
\label{app:string_validation}
To assess whether learned gene-pair relationships have support in an external biological resource, we examined kidney endothelial cells from the same Kidney Atlas used in the interpretability analyses and compared the top 50 genes paired with KDR by CellMSA with the human functional-association network in STRING v12 \cite{szklarczyk2023string}. Among these 50 genes, 26 have documented associations with KDR in STRING, corresponding to 52\% of the selected genes (Table~\ref{tab:string_validation}).

\begin{table}[htbp]
\centering
\caption{Descriptive overlap between CellMSA's top-ranked KDR gene pairs in kidney endothelial cells and STRING v12 functional associations.}
\label{tab:string_validation}
\begin{tabular}{lccc}
\toprule
Query gene & Top-ranked partners & STRING-supported partners & Supported fraction \\
\midrule
KDR & 50 & 26 & 52\% \\
\bottomrule
\end{tabular}
\end{table}

This overlap provides external database support for a subset of the learned associations. We report it as a descriptive comparison, without claiming statistically significant enrichment over a random background. The gene-pair representations are learned from expression data and should be interpreted as context-dependent statistical dependencies and potential functional associations, rather than direct evidence of causal gene-regulatory relationships.
\FloatBarrier

\subsection{Ablation study of pretraining objectives}
\label{app:ablation_objectives}
Masked gene modeling $\mathcal{L}_{\mathrm{MGM}}$ serves as the basic self-supervised objective for single-cell representation learning. 
Beyond this local reconstruction objective, we further introduce cell-level objectives to encourage the model to produce informative global cell embeddings. 
Specifically, $\mathcal{L}_{\mathrm{CCE}}$ uses cell-type metadata as weak supervision, guiding the representation to focus on functionally relevant biological patterns and improving robustness to batch effects. 
Meanwhile, $\mathcal{L}_{\mathrm{Rec}}$ encourages the cell embedding to preserve the unique transcriptomic profile of each cell, preventing the representation from collapsing to cell-type-level information only.

\begin{figure}[ht]
    \centering

    \begin{subfigure}[t]{0.32\linewidth}
        \centering
        \includegraphics[width=\linewidth]{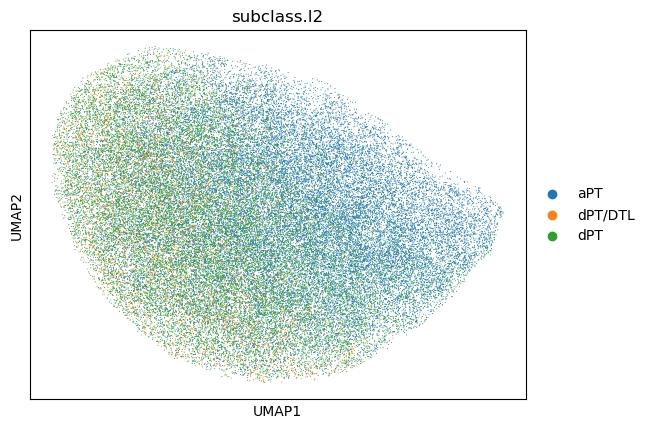}
        \caption{$\lambda_1=0$ and $\lambda_2=1$}
        \label{fig:b3-1}
    \end{subfigure}
    \hfill
    \begin{subfigure}[t]{0.32\linewidth}
        \centering
        \includegraphics[width=\linewidth]{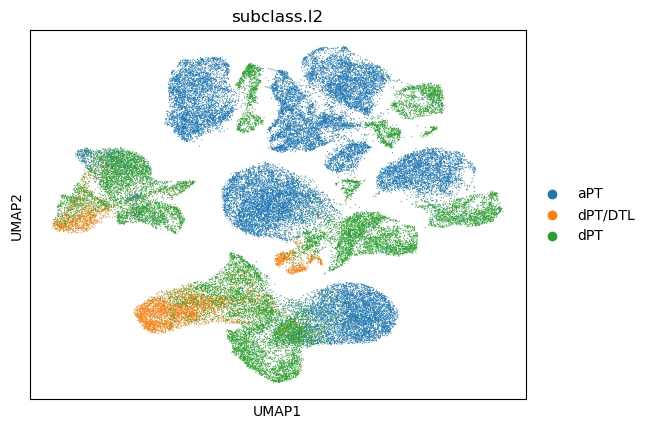}
        \caption{$\lambda_1=1$ and $\lambda_2=0$}
        \label{fig:b3-2}
    \end{subfigure}
    \hfill
    \begin{subfigure}[t]{0.32\linewidth}
        \centering
        \includegraphics[width=\linewidth]{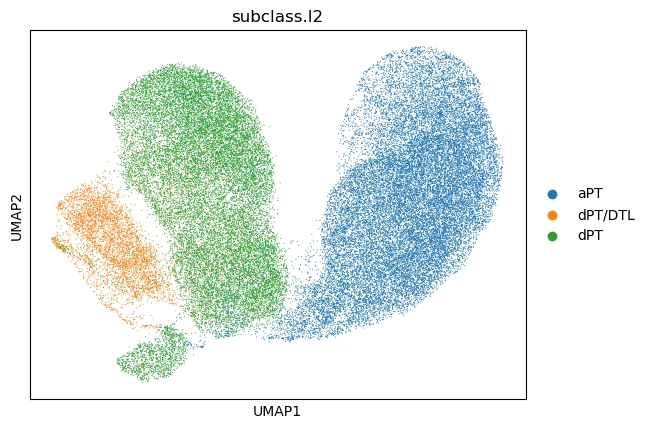}
        \caption{$\lambda_1=0.01$ and $\lambda_2=10$}
        \label{fig:b3-3}
    \end{subfigure}

    \caption{Visualization results across three PT cell states in different pretraining objective settings.}
    \label{fig:b3}
\end{figure}

These two objectives play complementary but partially competing roles. As shown in Figure \ref{fig:b3}, we use the model’s UMAP \cite{mcinnes2018umap} visualization results across three cell states of PT cells to illustrate its effects (for details of the dataset, see Appendix~\ref{app:downstream_dataset}).
If $\mathcal{L}_{\mathrm{CCE}}$ dominates, the learned representation may overemphasize cell-type identity and discard biological variation beyond cell types, such as disease states or cellular activation programs (as Figure \ref{fig:b3-1}). 
Conversely, if $\mathcal{L}_{\mathrm{Rec}}$ dominates, the representation may retain excessive input-specific noise and become less robust to sequencing depth, dropout, and batch effects (as Figure \ref{fig:b3-2}). 
Based on preliminary experiments, we set $\lambda_1=0.01$ and $\lambda_2=10$, which provides a good empirical balance between preserving cell-specific expression information and learning batch-robust cell-type-aware representations (as Figure \ref{fig:b3-3}).

\subsection{Robustness to missing nonzero genes}
\label{app:gene_dropout}
We evaluate the stability of CellMSA representations under incomplete expression inputs using Tabula Sapiens-Blood. We randomly remove 10\%, 30\%, or 50\% of the nonzero genes in cells and calculate the Pearson correlation between the original and corrupted cell embeddings. Table~\ref{tab:gene_dropout} reports the mean and median correlations across the evaluated cells.

\begin{table}[htbp]
\centering
\caption{Embedding stability under random removal of nonzero genes on Tabula Sapiens-Blood. The reported values compare original and corrupted embeddings.}
\label{tab:gene_dropout}
\begin{tabular}{lccc}
\toprule
Proportion of nonzero genes removed & 10\% & 30\% & 50\% \\
\midrule
Mean Pearson correlation & 0.996 & 0.988 & 0.972 \\
Median Pearson correlation & 0.998 & 0.993 & 0.984 \\
\bottomrule
\end{tabular}
\end{table}

Even after removal of 50\% of nonzero genes, the mean and median correlations remain 0.972 and 0.984, respectively. These results support representation stability under the tested random gene-removal setting. They do not directly measure downstream prediction accuracy or establish robustness to every form of missing-gene or sparsity pattern.
\FloatBarrier

\section{Experiment settings for pretraining and downstream tasks} 
\label{app:experiment_settings}
\subsection{Pretraining settings}
\label{app:pretraining_setting}

\paragraph{Corpus composition.}
After excluding datasets used for downstream evaluation, the pretraining corpus contains approximately 109 million human cell observations, including 65.6 million primary observations identified by \texttt{is\_primary\_data}. Thus, 109 million denotes observations rather than a count of unique biological cells. The non-primary observations are retained; our corpus audit indicates that most originate from re-aggregations within the same study or atlas integrations across studies. A CellMSA training instance is a target-cell--context pair, so observations of the same cell in different datasets can have different batch and neighborhood contexts. This contextual distinction does not remove the possibility of overrepresenting particular cells or atlas-specific structures.

\begin{figure}[ht]
    \centering
    \includegraphics[width=1\linewidth]{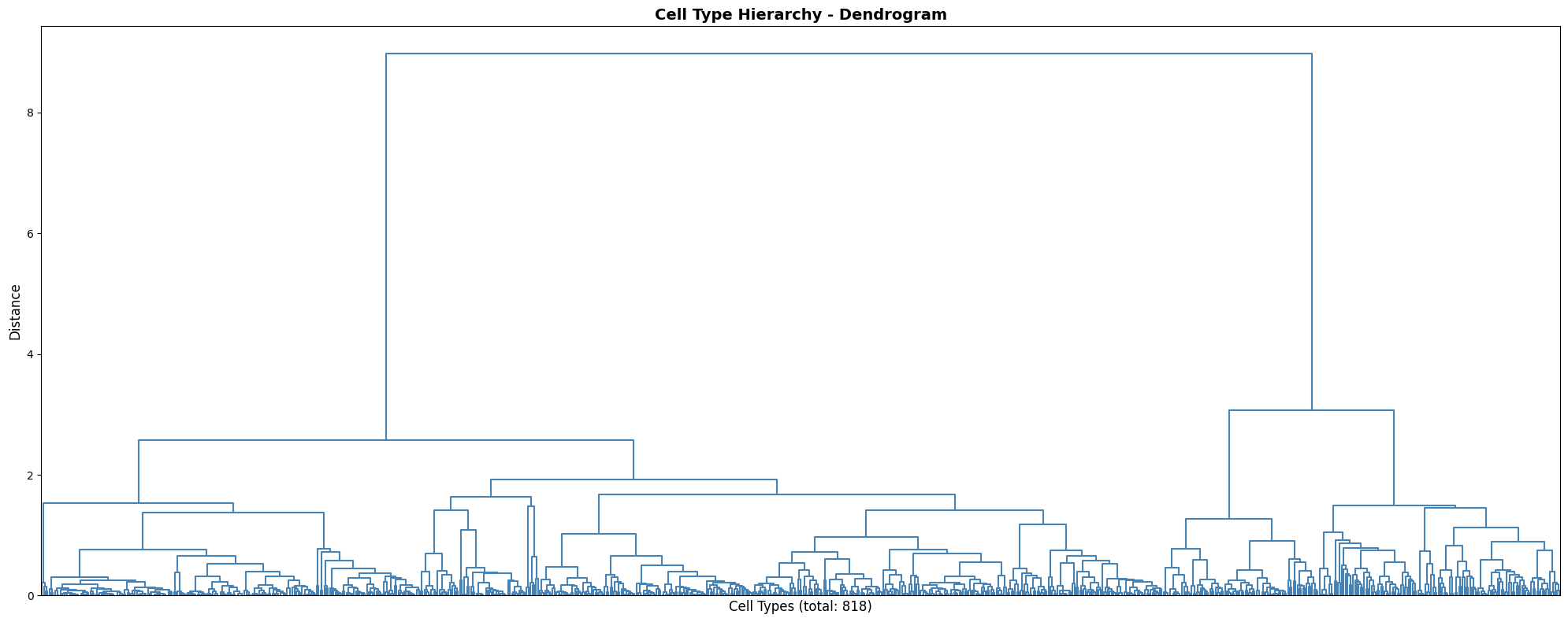}
    \caption{Dendrogram of the cell-type hierarchy.}
    \label{figc1}
\end{figure}

As part of pretraining data construction, we built a hierarchical clustering to define related-type neighbors for CellMSA sampling. For each cell type, we aggregated raw-count expression profiles across batches after total-count normalization to 10,000 counts and log1p transformation, producing a mean expression vector over 61,982 vocabulary genes. We then L2-normalized these vectors and computed a pairwise cosine similarity matrix, which was converted to cosine distance and used as input to Ward hierarchical clustering. The dendrogram was cut into 28 clusters, approximately following the square-root heuristic, $ k\approx \sqrt{818}$. For each cell type, we selected up to 50 related cell types from the same cluster, ranked by cosine similarity. The resulting hierarchy covered all 818 cell types, with cluster sizes ranging from 2 to 144. As a sanity check, within-cluster cosine similarity was substantially higher than between-cluster similarity on average, 0.845 versus 0.581. The constructed hierarchy is summarized by a dendrogram (Fig.~\ref{figc1}), and representative mappings from query cell types to their neighbors are reported to illustrate the related-type neighbor selection procedure (Table~\ref{tab:related-type-examples}). 

\begin{table}[t]
\centering
\caption{Representative related-type neighbors used for pretraining sampling.}
\label{tab:related-type-examples}
\resizebox{\textwidth}{!}{
\begin{tabular}{lccc}
\toprule
Query cell type & Cluster & \# related & Top related cell types (cosine similarity) \\
\midrule
CD4-positive, alpha-beta T cell & 2 & 50 & \makecell[l]{regulatory T cell (0.988);\\ CD4-positive helper T cell (0.986);\\ effector memory CD4-positive, alpha-beta T cell (0.986);\\ CD8-positive, alpha-beta T cell (0.985)} \\
\midrule
B cell & 2 & 50 & \makecell[l]{naive B cell (0.987);\\ memory B cell (0.984);\\ class switched memory B cell (0.982);\\ mature B cell (0.978)} \\
\midrule
fibroblast & 10 & 50 & \makecell[l]{thymic fibroblast type 2 (0.973);\\ fibroblast of breast (0.972);\\ stromal cell (0.972);\\ mesenchymal stem cell (0.966)} \\
\midrule
macrophage & 3 & 50 & \makecell[l]{monocyte (0.974);\\ Kupffer cell (0.969);\\ alternatively activated macrophage (0.969);\\ lung macrophage (0.968)} \\
\midrule
endothelial cell & 9 & 34 & \makecell[l]{endothelial cell of vascular tree (0.970);\\ retinal blood vessel endothelial cell (0.951);\\ endothelial cell of periportal hepatic sinusoid (0.938);\\ capillary endothelial cell (0.934)} \\
\bottomrule
\end{tabular}}
\end{table}

Pretraining is implemented based on the PyTorch framework. 
We use the AdamW optimizer \cite{kingma2014adam} with a schedule that first linearly warms up the learning rate and then keeps it constant at 1e-5. Pretraining is conducted on a server equipped with four NVIDIA Tesla A800 GPUs and  1 TB of memory, and takes approximately 20 days to complete. More detailed experimental configurations are summarized in Table~\ref{tab:cellmsa_hyperparameters}.

\begin{table}[ht]
    \caption{Pretraining Configurations of CellMSA}
    \label{tab:cellmsa_hyperparameters}
    \centering
    \resizebox{0.8\textwidth}{!} {
        \begin{tabular}{lll}
        \toprule
         & \textbf{Hyperparameter} & \textbf{Value} \\
        \midrule
        Input and context &
        {\makecell[l]{Gene vocab size \\ Max sequence length \\ Number of neighbors \\ Expression bins }} &
        {\makecell[l]{61,982 \\ 2,048 \\ 40 \\ 11}} \\
        \midrule
        Backbone &
        {\makecell[l]{Parameters \\ Hidden size \\ Pair embedding size \\ Cell embedding size}} &
        {\makecell[l]{47.13M \\ 512 \\ 8 \\ 512}} \\
        \midrule
        CellMSA-module &
        {\makecell[l]{Number of layers \\ MSA hidden size \\ Outer product hidden size \\ Attention heads \\ Attention head dim \\ Row dropout}} &
        {\makecell[l]{4 \\ 128 \\ 8 \\ 8 \\ 32 \\ 0.15}} \\
        \midrule
        GenePairformer &
        {\makecell[l]{Number of layers \\ Pair-biased attention heads \\ Attention head dim \\ Row dropout}} &
        {\makecell[l]{6 \\ 8 \\ 32 \\ 0.25}} \\
        \midrule
        Pretraining objectives &
        {\makecell[l]{Masked gene modeling probability \\ Mask replace probability \\ Random replace probability \\ Reconstruction probability \\ Reconstruction weight \\ CCE weight \\ CCE temperature \\ CCE interval}} &
        {\makecell[l]{0.15 \\ 0.8 \\ 0.1 \\ 0.1 \\ 0.01 \\ 10.0 \\ 0.05 \\ 1}} \\
        \midrule
        Pretraining &
        {\makecell[l]{Optimizer \\ Scheduler \\ Max learning rate \\ Weight decay \\ Warmup steps \\ Batch size per GPU \\ Number of GPUs \\ Gradient accumulation \\ Effective batch size \\ Mixed precision}} &
        {\makecell[l]{AdamW \\ Linear warmup then constant \\ 1e-5 \\ 0.05 \\ 1,000 \\ 4 \\ 4 \\ 2 \\ 32 \\ Enabled}} \\
        \bottomrule
        \end{tabular}
    }
\end{table}

\subsection{Downstream tasks datasets}
\label{app:downstream_dataset}

We assemble a set of benchmark datasets to evaluate the performance of CellMSA across various downstream tasks.
The following discussion will be structured according to the dataset of cells involved.

\paragraph{Tabula Sapiens}
The Tabula Sapiens dataset, sourced from the work \cite{Tabula_Sapiens}, is a comprehensive human single-cell transcriptomic atlas. It has uncovered the transcriptomic features of 475 distinct cell types by analyzing live cells from multiple human tissues. The data is derived from 59 meticulously selected samples, encompassing a broad range of tissue types from the bladder to the vasculature, involving donors of varying genders, ethnicities, and ages. The project has analyzed a total of 483,152 cells, including a substantial number of immune cells, epithelial cells, endothelial cells, and stromal cells.

\paragraph{Kidney Atlas.} The Kidney Atlas dataset comprises a comprehensive collection of human kidney single-cell profiles from 77 donors, including 26 healthy individuals, 14 patients with acute kidney injury (AKI), and 37 patients with chronic kidney disease (CKD)~\cite{de2021rationale}. The dataset contains annotations for 43 kidney cell types and captures a broad spectrum of disease-associated transcriptional variation. 
In this work, we focus on proximal tubule (PT) cells, because the proximal tubule compartment is strongly affected in both AKI and CKD. Within the PT compartment, we use three annotated cell states as classification labels: aPT, dPT, and dPT/DTL. These labels represent a progression from normal-like PT cells to injured PT cells and further to severely injured PT cells with descending thin limb-like transcriptional identity drift. This setting provides a challenging benchmark for evaluating whether learned representations can resolve fine-grained disease-associated cell states within the same epithelial lineage.

\paragraph{Replogle Dataset}
The Replogle dataset is a large-scale single-cell CRISPR perturbation dataset that measures transcriptomic responses to genetic perturbations across multiple human cell lines \cite{replogle2022mapping}. 
Following the perturbation prediction setting in our main experiments, we use four human cell lines and retain the top 100 perturbations that are present in all four cell lines and have the largest numbers of cells. 
Together with the control group, the filtered dataset contains approximately 132k cells. 
This dataset provides a challenging setting for evaluating whether learned cell representations can support accurate prediction of perturbation-induced transcriptional shifts across cellular contexts.

\subsection{Downstream tasks settings}
\label{app:downstream_setting}

\paragraph{Baselines}
For the large-scale single-cell foundation models, we utilized their official codebases and specific pretrained checkpoints to ensure strict reproducibility. Specifically, we employed the scGPT whole-human foundation model, the Geneformer V2 model with 104M parameters (geneformer\_V2\_104M), the State SE-600M checkpoint, and the Stack-Large model. We additionally use the official CellPLM implementation and pretrained checkpoint for the two classification tasks. All foundation models were executed in a zero-shot manner using their default hyperparameter configurations for embedding extraction, without any task-specific fine-tuning.

\paragraph{Batch Integration.}
We utilized the Tabula Sapiens dataset, accessed via the CELLxGENE Census portal (dataset id: 53d208b0-2cfd-4366-9866-c3c6114081bc). 
All batch integration metrics were calculated utilizing the scib-metrics package (v0.5.9). For all comparative analyses, the PCA embedding was designated as the pre-integrated baseline to anchor relative metric calculations. donor\_id and cell\_type served as the batch and biological label keys, respectively. 
For the main \emph{label-informed integration} setting, we construct CellMSA and Stack contexts using the cell-type annotations provided by Tabula Sapiens. This setting targets atlas alignment with available coarse annotations. Because the other baselines do not use these labels for representation extraction, the main benchmark includes methods with different information access. Appendix~\ref{app:label_free_integration} therefore additionally evaluates Bladder under a label-free protocol: context neighbors are retrieved with the HVG--PCA--KNN strategy used for classification, without cell-type labels. Evaluation labels are retained only for computing the biological-conservation metrics.

Specifically, for clustering-dependent metrics (NMI and ARI), we utilized the Leiden algorithm \cite{traag2019louvain}. The isolated label metric was omitted from this evaluation pipeline according to the tutorial of scib-metrics. We evaluated the integration performance independently on each tissue and reported the macro-averaged scores. Detailed metric breakdowns for individual tissues are provided in Appendix \ref{app_metric_batch}.

\paragraph{Cell Type and Cell State Classification.}
For fine-grained classification, we evaluate CellMSA on two representative settings: cell type annotation on Tabula Sapiens-Blood and PT cell state classification on the Kidney Atlas. 
For the cell type annotation task, we selected cell types representing at least 1\% of the total cell count. For the PT cell state classification task, we restrict the dataset to proximal tubule (PT) cells from healthy, AKI, and CKD donors, and classify cells into three PT states: aPT, dPT, and dPT/DTL. This task is designed to evaluate whether a representation can distinguish normal, injured, and severely injured PT states within the same kidney epithelial lineage.
For each dataset, we split the cells by donor ID into training/validation/test sets with a ratio of 7:1:2. We conduct five random experiments and report the mean results and standard deviations.
To avoid label leakage during context construction, we do not use the ground-truth cell type/state labels of validation or test cells to retrieve their contextual neighbors. Instead, we first compute highly variable genes (HVGs), perform principal component analysis (PCA), and then use K-nearest-neighbor (KNN) search in the PCA space to select the type of contextual cells for validation and test samples. 
After obtaining cell representations from the pretrained model, we train a three-layer MLP classification head on top of the cell embeddings. We evaluate classification performance using accuracy, macro-F1 score, and weighted-F1 score.

\paragraph{Perturbation Prediction.}
For perturbation prediction, we follow the experimental setting of STATE and evaluate whether the learned cell representations can support accurate prediction of perturbation-induced transcriptional changes. We use 45\% of the perturbations in the HepG2 cell line as the test set and 5\% as the validation set. The remaining 50\% of HepG2 perturbations, together with all perturbations from the K562, Jurkat, and RPE1 cell lines, are used for training. 

For this task, we use a task-specific definition of the MSA-like context. Because this task focuses on transcriptomic changes induced by perturbations, we construct context cells within the same cell line. Cells with the same perturbation are used as same-batch context $\mathcal{N}^{\mathrm{same~batch}}$, whereas cells from the same cell line but with different perturbations are treated as cross-type context $\mathcal{N}^{\mathrm{cross~type}}$. Here, ``cross-type'' refers to perturbation-defined cellular states rather than ontology-level cell types, enabling the model to compare transcriptional differences across perturbation conditions under a shared cellular background. In addition, to prevent data leakage, we do not use the post-perturbation expression data from the test/validation sets as the cellular context.

Following STATE, we use its open-source implementation to train a STATE-ST model on top of the cell representations produced by each encoder. The resulting perturbation predictions are evaluated using the Cell-Eval package. We use Pearson $\Delta$ as the primary metric, since it directly measures the agreement between predicted and true perturbation effects. In addition, we report three representative metrics related to differentially expressed genes: PRAUC, DE Overlap, and Spearman-FC. The detailed definitions and computation procedures of these metrics are provided in Appendix~\ref{app_metric_perturb}. The experiments require training STATE-ST for each run, which is computationally too expensive; therefore, we only use a fixed split for a single experiment.

\subsection{Metrics of batch integration}
\label{app_metric_batch}
We adopt the widely recognized Single-cell Integration Benchmarking (scIB) framework \cite{scib}. scIB groups the metrics into two broad categories: (1) conservation of biological variance and (2) removal of batch effects. The latter category is further divided into conservation of variance from cell identity labels, and conservation of variance beyond cell identity labels. Scores from the first category include NMI, ARI, ASW and cLISI. Scores from the second category include BRAS, iLISI, kBET, Graph connectivity and PCR (Principal component regression). 

\textbf{NMI}: Normalized Mutual Information (NMI) quantifies the overlap between ground-truth cell-type labels and unsupervised clustering results, scaled to a 0–1 range where 1 indicates perfect agreement. We optimized Leiden clustering resolution to maximize NMI for evaluation.

\textbf{ARI}: Adjusted Rand Index (ARI) \cite{hubert1985comparing} measures the similarity between two clusterings by correcting the Rand Index for chance, yielding scores from -1 to 1 with 1 representing perfect match. The evaluation used Leiden clusters optimized for the highest NMI on the integrated data.

\textbf{ASW}: Average Silhouette Width (ASW) \cite{rousseeuw1987silhouettes} captures cell-type separation. Higher ASW indicates better bio-conservation.

\textbf{BRAS}: Batch-removal-adapted Silhouette (BRAS) \cite{rautenstrauch2025shortcomings} modifies the silhouette score by taking 1 minus the absolute silhouette for each cell within a label and macro-averaging over all cell types. It produces a score between 0 and 1, with higher values indicating superior batch mixing.

\textbf{Graph LISI}: Graph LISI \cite{korsunsky2019fast} evaluates batch mixing (\textbf{iLISI}) and cell-type separation (\textbf{cLISI}) using neighborhoods on an integrated kNN graph with graph distances, rescaled to 0–1. Higher iLISI corresponds to better mixing, and higher cLISI to better cell-type separation.

\textbf{kBET}: kBET \cite{buttner2019test} tests whether local batch composition matches the global distribution using a kNN neighborhood test, with the rejection rate averaged per cell type and subtracted from 1. A final score close to 1 indicates effective batch removal.

\textbf{Conn}: Graph connectivity (Conn) \cite{scib} measures the fraction of cells of the same label that belong to the largest connected component in the integrated kNN graph, ranging from 0 to 1. A score of 1 means all cells sharing a label are fully connected.

\textbf{PCR}: Principal component regression (PCR) \cite{scib} quantifies batch effects by summing, across all PCs, the variance explained multiplied by the R² from regressing the batch variable onto each PC. We report the normalized PCR comparison score used by scib-metrics, where higher values indicate better batch removal.

The overall score, $S_{overall,i}$, for each integration run i was calculated by taking the weighted mean of the batch removal score, $S_{batch,i}$, and the bio-conservation score, $S_{bio,i}$, following the equation \cite{scib}:

$S_{{\mathrm{overall}},i} = 0.6 \times S_{{\mathrm{bio}}} + 0.4 \times S_{{\mathrm{batch}}}.$

In turn, these partial scores were computed by averaging all metrics that contribute to each score.

\subsection{Metrics of perturbation prediction}
\label{app_metric_perturb}

We evaluate perturbation prediction using metrics from Cell-Eval \cite{adduri2025predicting}. 
For each perturbation, we first compute the mean perturbation effect relative to control cells:
\[
\Delta^{\mathrm{real}}_p = \bar{x}^{\mathrm{real}}_p - \bar{x}^{\mathrm{real}}_{\mathrm{ctrl}}, 
\qquad
\Delta^{\mathrm{pred}}_p = \bar{x}^{\mathrm{pred}}_p - \bar{x}^{\mathrm{pred}}_{\mathrm{ctrl}} .
\]
\textbf{Pearson $\Delta$} is then computed as the Pearson correlation between 
$\Delta^{\mathrm{real}}_p$ and $\Delta^{\mathrm{pred}}_p$ across genes. 
This metric directly measures whether the predicted transcriptional shift matches the true perturbation effect.

We also report three representative metrics related to differentially expressed (DE) genes:

\textbf{PRAUC} evaluates whether the model can recover truly significant DE genes by treating real DE significance as binary labels and using the predicted DE significance scores for ranking. 

\textbf{DE Overlap} measures the overlap between the top-ranked DE genes in the real and predicted perturbation profiles, sorted by absolute fold change. 

\textbf{Spearman-FC} computes the Spearman correlation between the real and predicted fold changes on real significant DE genes. 

Together, these metrics assess not only the overall perturbation effect but also whether the model correctly identifies and ranks perturbation-associated DE genes.

\section{Broader impacts}
\label{app:broader_impacts}

CellMSA may facilitate single-cell data analysis, disease-state characterization, and perturbation response prediction, thereby supporting biomedical discovery and therapeutic research. 
However, its predictions may be affected by dataset bias, batch effects, annotation noise, and incomplete cell-state coverage, and should therefore be regarded as hypothesis-generating rather than clinically actionable evidence. 
Upon public release, CellMSA should be restricted to research purposes, and any biomedical or translational application should require independent validation, biological experiments, and expert inspection.

\end{document}